\documentclass[twocolumn]{aastex701}

\usepackage{multirow}

\newcommand{\Msun}{M$_{\odot}$}

\newcommand{\Msunyr}{M$_{\odot}$~yr$^{-1}$}
\usepackage{xcolor}

\newcommand{\ralgn}[1]{\hfill #1} 

\begin{document}

\title{Dynamical formation of long-period exoplanets systems in evolving binary stars}

\author[orcid=0000-0001-6013-1772,sname='Isabel Lima']{Isabel~J.~Lima}
\affiliation{São Paulo State University (UNESP), School of Engineering and Sciences, Guaratinguet\'{a}, Av. Dr. Ariberto Pereira da Cunha, 333 - Pedregulho - Guaratinguet\'a, 12516-410, SP,  Brazil}
\affiliation{Laboratório Nacional de Astrofísica (LNA), Rua dos Estados Unidos 154, Bairro das Nações, Itajubá, 37504-364, MG, Brazil}
\email[show]{isabel.jesus@unesp.br}  

\author[orcid=0000-0002-1666-5141, sname='Rafael~Sousa']{Rafael~R.~Sousa} 
\affiliation{São Paulo State University (UNESP), School of Engineering and Sciences, Guaratinguet\'{a}, Av. Dr. Ariberto Pereira da Cunha, 333 - Pedregulho - Guaratinguet\'a, 12516-410, SP, Brazil}
\email{r.sousa@unesp.br}

\author[orcid=0000-0002-3949-6045
, sname='Silvia~Giuliatti~Winter']{Silvia~M.~Giuliatti~Winter} 
\affiliation{São Paulo State University (UNESP), School of Engineering and Sciences, Guaratinguet\'{a}, Av. Dr. Ariberto Pereira da Cunha, 333 - Pedregulho - Guaratinguet\'a, 12516-410, SP, Brazil}
\email{giuliatti.winter@unesp.br}

\author[orcid=0000-0003-1535-0866
, sname=' Diogo~Belloni']{ Diogo~Belloni} 
\affiliation{International Centre of Supernovae (ICESUN), Yunnan Key Laboratory of Supernova Research, Yunnan Observatories, CAS, Kunming 650216, China}
\email{diogo@ynao.ac.cn}

\author[orcid=0000-0002-7589-0998
, sname='Ernesto~Vieira']{Ernesto~Vieira} 
\affiliation{São Paulo State University (UNESP), School of Engineering and Sciences, Guaratinguet\'{a}, Av. Dr. Ariberto Pereira da Cunha, 333 - Pedregulho - Guaratinguet\'a, 12516-410, SP, Brazil}
\email{ernesto.vieira@unesp.br}

\author[orcid=0000-0001-5198-3025
, sname='Rosana~Araújo']{Rosana~A.~N.~Ara\'{u}jo} 
\affiliation{São Paulo State University (UNESP), School of Engineering and Sciences, Guaratinguet\'{a}, Av. Dr. Ariberto Pereira da Cunha, 333 - Pedregulho - Guaratinguet\'a, 12516-410, SP, Brazil}
\email{rosana.araujo@unesp.br}

\author[orcid=0000-0002-5084-168X, sname='Eder Martioli']{Eder Martioli} 
\affiliation{Laboratório Nacional de Astrofísica (LNA), Rua dos Estados Unidos 154, Bairro das Nações, Itajubá, 37504-364, MG, Brazil}
\email{emartioli@lna.br}

\begin{abstract}

The dynamical formation and evolution of long-period giant exoplanets have not been well constrained due to the few observational parameters. In this study, we explore the dynamical effects in a multi-planetary system around one star of an evolving wide binary system. We used an interpolated result of a \texttt{MESA} stellar evolution inside \texttt{REBOUND} $N$-body integrations to perform simulations with a range of distinct planetary masses in a circumstellar configuration \added{(S-type orbit)}, centered on a primary star that evolves from the main sequence star to a white dwarf. Control simulations without stellar evolution were performed to isolate the effects of mass loss. Although few exoplanets are currently known to possess long-period and moderate eccentricities, we investigated the evolution mechanisms of long-period gas giants within this specific regime. Although these exoplanets often remain undetected due to their wide orbits and long periods, we propose that their formation pathways are robust throughout the evolution of binary systems. \added{We also simulated an evolved single star in a multi-planetary system and concluded that the secondary star made the exoplanets more unstable and concentrated the survivors within a semi-major axis of 50~au.}

\end{abstract}



\section{Introduction}

The majority of stars in the solar neighborhood belong to binary or higher-order systems \citep{Duquennoy_1991, Raghavan_2010, Moe_2017},  which many of them host exoplanets\footnote{see the NASA Exoplanet Archive at \url{exoplanet.eu}.}.
Therefore, the study of exoplanets in the context of binaries stars is crucially important, given that binarity should be a relatively common environment in the universe. Recent studies have shown that the occurrence of exoplanets in binaries stars is comparable to the single stars. For example, \cite{Hirsch_2021} considered a volume-limited sample of $<25$~pc and an unbiased sample of single and multiple stellar systems. They performed a relative velocity (RV) search for giant exoplanets ($>0.1$~M$_{Jup}$) and found that the occurrence rate of exoplanets with masses between  ${0.1-10}$~M$_{\text{Jup}}$ in binaries is about 12\%, which is very similar to the occurrence for single stars (18\%). Wide binaries (semi-major axis $\geq100$~au) present a planet-occurrence rate more similar to that of single stars.

Theoretical studies show that planet-formation process is influenced by binarity across the evolutionary stages; for instance, the protoplanetary disk in these binary systems are statistically less frequent and less massive than those around single stars (for a review see 
\citealt{Thebault_2015_book}). Furthermore, the disk can be tidally truncated by a stellar companion, especially in wide binaries, which could potentially lead to the formation of more massive planets via fragmentation in a gravitationally unstable disk \citep{Fontanive_2021}. 
Many models predict that the effect of the secondary star depends on the orbital parameters of the  binary, specifically the semi-major axis ($a$), eccentricity ($e$), and inclination ($i$). The formation of a planet could be hindered by close, eccentric, or highly inclined companions \citep{Holman_1999, Thebault_2015_book, Cadman_2022}.
In late stages, assuming that the first stages of planet formation were successful, binaries with moderate to large perihelia (generally $q>100$~au) 
and with giant planets on low eccentricity orbits can also produce Earth-like planets \citep{Haghighipour_2007}.

\added{In terms of orbital mechanics, there are two possible types of orbit for exoplanets in binary systems (not including positions near the Lagrangian points L4; \citealt{Dvorak_1982}): a P-type orbit, in which the exoplanet orbits both binary components and S-type orbit, in which the exoplanet only orbits one of the binary components, while the second component acts as a perturbator \citep{Cuntz_2014}.}

Long-period exoplanets, defined here as those with orbital periods corresponding to  $a\geq5$~au, represent a historically under-sampled population in transit surveys. 
These planets are significantly constrained by severe observational biases inherent in current detection methodologies.  
Traditional indirect techniques, such as RV and transit photometry, are inherently biased toward short-period objects due to the requirement of long-term observational baselines$-$often exceeding several decades for a single orbital completion at these distances$-$and the diminishing signal amplitudes \citep[e.g.,][]{Scharf_2009}. The number of long-period planets that exhibit transits is significantly short. The probability of an eclipse due to the geometric configuration of the orbital planet's orientation relative to our line of sight through the system is proportional to $R_{s}/a$, where $R_{s}$ is the radius of the star and $a$ is the semi-major axis of the planet orbit. In other words, the probability decreases with the orbital period in the ratio of $1/P^{2/3}$ for long-period planets. Consequently, the majority of known exoplanets beyond $15$~au have been identified via direct imaging, a method that introduces its own selection effects by favoring young, self-luminous, and massive gas giants orbiting at wide separations from their host stars. 

\cite{Foreman_2016}, combining the measured detection efficiency and the catalog of exoplanet candidates, estimated the occurrence rate of long-period exoplanets orbiting between ${3-20}~$au and a radius ranging from ${0.1-1}$~R$_{\text{Jup}}$ to be around 2~$\pm$~0.7 planets per Sun-like star. Meanwhile, \cite{Bryan_2016} reported an occurrence rate of companions in exoplanet systems with one or two planets over ${1-20}$~M$_{\text{Jup}}$ and ${5-20}~$au of 52~$\pm$~5\% from a sample of 123 known exoplanet systems detected using the RV method. The authors also found that the eccentricity distribution was significantly higher in multi-body systems than in single-planet systems with no outer bodies. These results implied that long-period companions to hot Jupiters are common.

The dynamical evolution of planetary systems orbiting evolved single stars has been extensively studied over the past decades (e.g., \citealt{Rasio_1996}; \citealt{Villaver_2007}; \citealt{Mustill_2012}; \added{\citealt{Kratter_2012};} \citealt{Veras_2013}; \citealt{Mustill_2014}; \citealt{Veras_2016}; \citealt{Rao_2018}; \added{\citealt{Grunblatt_2018}; \citealt{Stephan_2018};} \citealt{Ronco_2020}; \added{\citealt{Stephan_2020}}; \citealt{Mustill_2024}; \citealt{Mauch-Soriano_2026}). Most of these studies adopt simplified prescriptions for stellar evolution, commonly relying on stellar evolutionary tracks from the Single Stellar Evolution (\texttt{SSE}) code \citep{Hurley_2000}, which provides a parametrized, yet computationally efficient, description of stellar mass and radius evolution. The problem becomes considerably more complex in the context of multiple stellar systems. In this regime, \cite{Hamers_2021} introduced a population synthesis framework for Multiple Stellar Evolution (\texttt{MSE}) that includes planetary systems and extending the \texttt{SSE} code approaches to hierarchical systems. In that framework, however, binary evolution$-$particularly during phases of mass transfer$-$is still treated in a simplified manner, largely following the prescriptions implemented in the Binary Stellar Evolution (\texttt{BSE}) code \citep{Hurley_2002}.

More recently, \cite{Baronett_2022} developed a flexible and machine-independent framework to incorporate stellar evolution into $N$-body simulations through \texttt{REBOUNDX} \citep{Tamayo_2020}. In order to investigate the impact of the Sun's post-main-sequence evolution on the outer giant planets, the authors used the code Modules for Experiments in Stellar Astrophysics (\texttt{MESA}, \citealt[][r15140]{Paxton2011,Paxton2013,Paxton2015,Paxton2018,Paxton2019,Jermyn2023}), which stellar parameter data were interpolated and implemented with a constant time-lag tidal model without evolving stellar spins. This approach enables stellar evolution outputs from external codes to be seamlessly coupled to \texttt{REBOUND} \citep{Rein2012, Rein2015Spiegel, Rein2015Tamayo} integrations.

Expanding on this, \cite{Xing_2025} focused on accurately modeling the binary stellar evolution of systems hosting circumbinary planets, with particular emphasis on the mass transfer phase. In their work, the authors coupled the state-of-the-art stellar evolution code \texttt{MESA} with the high-performance $N$-body integrator \texttt{REBOUND} developing the $N$-Body Binary Stellar Evolution (\texttt{NBSE}) code. Unlike \texttt{BSE}-like approaches, in which mass transfer rates are obtained through parametrized prescriptions, \texttt{MESA} enables a self-consistent calculation of mass transfer, accounting for stellar rotation, tidal interactions, mass, and angular momentum losses through stellar winds and non-conservative mass transfer throughout the binary evolution. \texttt{NBSE} code allows for a fully self-consistent treatment of both single and binary stellar evolution while simultaneously tracking the dynamical response of circumbinary planets by resolving the detailed physics of stellar interactions and their feedback on planetary orbits. \texttt{NBSE} code represents a significant advance over simplified population synthesis approaches in the study of the coupled evolution of binaries and their planetary systems.

In this paper, we investigate the dynamical formation of giant planets in systems with evolving binary stars in \added{S-type orbit} 
using \texttt{MESA} and \texttt{REBOUND}.
We consider the multi-planetary system after the gaseous protoplanetary disk has dissipated, when the remaining planets undergo dynamical evolution that includes collisions, mergers, scattering, and significant ejection of planets from the system over tens to hundreds of millions of years (Stage~2; see details in \citealt{Juric_2008}). We also include control-run simulations of the multi-planetary systems without stellar evolution in order to compare and isolate the effects of stellar evolution.

\section{MESA model assumptions}
\label{Sec:MESAAssumptions}

We used the \texttt{MESA} code to carry out the evolution of zero-age main-sequence stars (ZAMS) to a wide binary composed by a white dwarf (primary) and a main-sequence star (secondary). We describe here our assumptions for single stars and binary evolution, which are summarized in Table~\ref{Tab:MESAAssumptions}.
For reference, our approach follows closely that by \citet{Belloni_2024c}\,\footnote{\added{See \cite{ZENODO.10841636} in} \href{https://zenodo.org/records/10841636}{https://zenodo.org/records/10841636}} and \citet{Belloni_2024d}\,\footnote{\added{See  \cite{ZENODO.10937460} in} \href{https://zenodo.org/records/10937460}{https://zenodo.org/records/10937460}}.

The \texttt{MESA} equation of state is a blend of the OPAL \citep{Rogers2002}, SCVH \citep{Saumon1995}, FreeEOS \citep{Irwin2004}, HELM \citep{Timmes2000}, PC \citep{Potekhin2010} and Skye \citep{Jermyn2021} equations of state. Nuclear reaction rates are a combination of rates from NACRE \citep{Angulo1999}, JINA REACLIB \citep{Cyburt2010}, plus additional tabulated weak reaction rates \citep{Fuller1985,Oda1994,Langanke2000}. Screening is included via the prescription of \citet{Chugunov2007} and thermal neutrino loss rates are from \citet{Itoh1996}. Electron conduction opacities are from \citet{Cassisi2007} and radiative opacities are primarily from OPAL \citep{Iglesias1993,Iglesias1996}, with high-temperature Compton-scattering dominated regime calculated using the equations of \citet{Buchler1976}.

\begin{deluxetable*}{cc}
\digitalasset
\tablewidth{0pt}
\tablecaption{Adopted stellar and binary evolution parameters in \texttt{MESA} code. \label{Tab:MESAAssumptions}}
\tablehead{
\colhead{Parameter} & \colhead{Values } 
} 
\startdata
Initial separation &  $100$~au \\
\added{Final separation} &  \added{$202$~au}\\
Initial WD progenitor mass &  $2$~\Msun \\
Initial companion mass &  $0.8$~\Msun \\
Metallicity Z  &  $0.02$ \\ 
Reimers's parameter &  0.1$^{a}$ \\ 
Bl\"ocker's parameter &  0.02$^{b}$ \\ 
Mixing length & $2.0\,H_p$$^{c}$ \\
\vspace{-0.1cm}
Extent of diffusive & \multirow{2}{*}{$0.016\,H_p$}\\
exponential core overshooting &  \\
\vspace{-0.1cm}
Criterion for stability & \multirow{2}{*}{Schwarzschild $(\nabla_{\rm ad} = \nabla_{\rm rad})$} \\
against convection & \\
Nuclear network & \texttt{auto$\_$extend$\_$net} \\
\enddata
\tablecomments{ $^{a}$ Wind prescription (before AGB) is based on \citet{Reimers_1975}. $^{b}$ Wind prescription (AGB) is based on \citet{Blocker_1995}. $^{c}$ MLT is based on \citet{Henyey_1965}.}
\end{deluxetable*}

We adopted a metallicity of ${Z=0.02}$ and assumed the grey Eddington T($\tau$) relation to calculate the outer boundary conditions of the atmosphere, using a uniform opacity that is iterated to be consistent with the final surface temperature and pressure at the base of the atmosphere \citet{Ferguson2005}.
For the evolutionary phases during the core is convective, that is, core hydrogen and helium burning, we took into account exponential diffusive overshooting, assuming a smooth transition in the range ${1.2-2.0}$~\Msun~\citep[e.g.,][]{Anders_2023}. We assumed that the extent of the overshoot region corresponds to ${0.016~H_{\rm p}}$ \citep[e.g.,][]{Schaller_1992,Freytag_1996,Herwig_2000}, with $H_{\rm p}$ being the pressure scale height at the convective boundary. We included rotation in our calculations and we set the rotation of the ZAMS to $1$\% of their critical rotation rates. We treate convective regions using the scheme by \citet{Henyey_1965} for the mixing-length theory, assuming that the mixing length is ${2\,H_{\rm p}}$ \citep[e.g.,][]{Joyce_2023}. We also included mixing of angular momentum and rotationally induced mixing processes (Solberg-Hoiland, secular shear instability, Eddington-Sweet circulation, Goldreich-Schubert-Fricke and, Spruit-Tayler dynamo), which in \texttt{MESA} are treated following \citet{Heger_2000} and \citet{Heger_2005}.

\begin{figure}[tbh!]
    \includegraphics[width=\columnwidth]{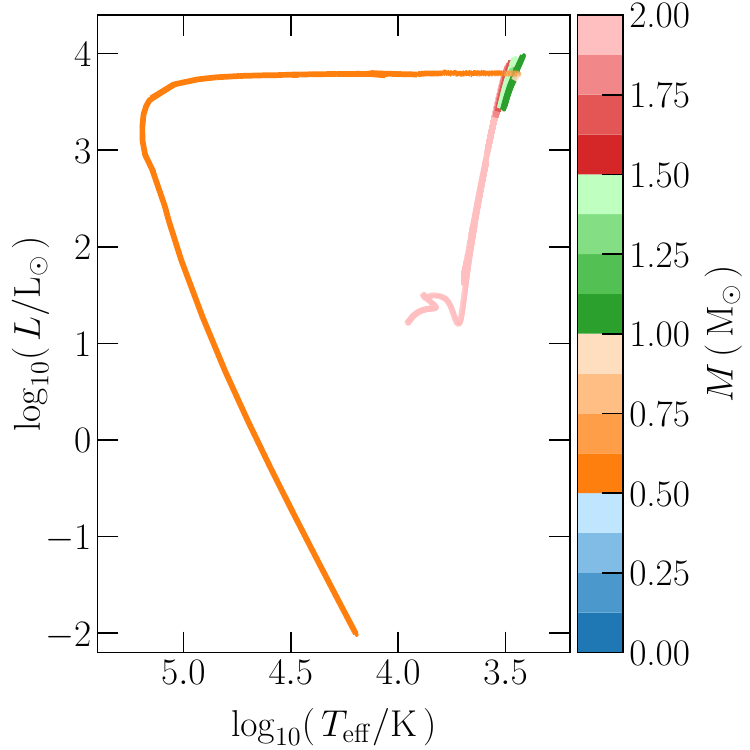}
 \caption{Evolution of the primary star obtained with the \texttt{MESA} code from the ZAMS to the white dwarf in the HR diagram. The track is color coded according to the mass. The mass loss starts being significant during the AGB phase, dropping to ${\sim0.58}$\,M$_\odot$. The orange line indicates the post-AGB phase to white dwarf.}
    \label{fig:HRdiagram}
\end{figure}

We allowed the stars to experience mass loss through winds, applying the \citet{Reimers_1975} model during pre-asymptotic giant branch (AGB) evolution with a wind efficiency of $0.1$, which is consistent with the low efficiency inferred from observations of solar-metallicity open clusters \citep{Miglio_2012,Handberg_2017}. During AGB evolution, we adopted the prescription proposed by \citet{Blocker_1995}, setting the wind efficiency to $0.02$, which is supported by the calibration performed by \citet{Ventura_2000} using the luminosity function of lithium-rich stars in the Magellanic Clouds. For the nuclear network, we assumed the auto-extended scheme, which automatically extends the net as needed.
Although we included wind mass loss and nuclear burning in our simulations for almost the entire evolution, we turn them off when the envelope mass becomes sufficiently small (${<0.001}$~\Msun) after the AGB phase, since beyond this point shell burning and winds are negligible. Figure~\ref{fig:HRdiagram} shows the evolutionary track of the primary star from ZAMS to the white dwarf in the Hertzprung-Russel (HR) diagram. 

We allowed both stars in the binary to eventually synchronize with the orbit due to tidal interaction. The Roche-lobe radius of each star was computed using the fit of \citet{Eggleton1983}. Regarding wind accretion, we adopted the Bondi-Hoyle-Lyttleton prescription \citep{Hoyle_1939,Bondi_1944}.
Following \citet{Hurley_2002}, we enforced that the main-sequence accretor ($M_{\rm a}$) cannot accrete the mass being transferred from the AGB donor ($M_{\rm d}$) at a rate higher than allowed by its thermal timescale ($\tau_{\rm KH,a}$).
We assumed that ${\dot{M}_{\rm a} = \eta\,\dot{M}_{\rm d}}$, where ${\eta=M_{\rm a}/\tau_{\rm KH,a}}$, if ${\dot{M}_{\rm d}>M_{\rm a}/\tau_{\rm KH,a}}$, or ${\eta=1}$, otherwise.
The non-accreted material is assumed to be lost from the vicinity of the main-sequence accretor as fast wind.
This critical rate is typically on the order of a few ${10^{-7}}$~\Msunyr.

\section{Numerical Simulations}
\label{Sec:NumericalS}

We used \texttt{REBOUND} $N$-body integrator to model the dynamical evolution of different planetary systems under the gravitational influence of evolving binary stars. Simulations were performed using the \texttt{IAS15} adaptive high-order integrator \citep{Rein2015Spiegel}, with an initial timestep of 0.5~yr up to 1.4~Gyr. We performed simulations over the entire lifetime of the primary star, rather than focusing only on the short period of significant mass loss. Since both stars lose mass very slowly, this produces secular effects on the orbits of the exoplanets. In compact orbital configurations, these effects can trigger instabilities, as shown in simulations by \cite{Veras_2017}. Collisions were treated as perfectly inelastic, resulting in mergers that conserve total linear momentum. Planets were removed from the simulation if their distance from the primary star dropped below 1~au or if their orbital eccentricity exceeded unity.

\subsection{Simulation Setup: Evolving Binary Stars Only}

The binary evolution has a timestep longer than the planet's dynamical evolution requires. To synchronize the $N$-body simulation with the precalculated binary evolution model obtained from \texttt{MESA}, we linearly interpolated the time series as a function of the stellar parameters such as masses and radii as obtained from \texttt{MESA}'s output. A similar approach was used by \cite{Baronett_2022} and  \cite{Xing_2025}, who also interpolated the \texttt{MESA} data to be used in \texttt{REBOUND}.
\added{Therefore, we obtained from the MESA code the stellar evolution parameters, such as the masses and radii of the primary and secondary stars. We set the secondary's eccentricity and inclination to zero. In \texttt{REBOUND} code, we considered the secondary eccentricity  initially to be equal to zero, while allowing the inclination to vary freely from 0.001 onward.}

\begin{figure}[tbh!]
\includegraphics[width=\columnwidth]{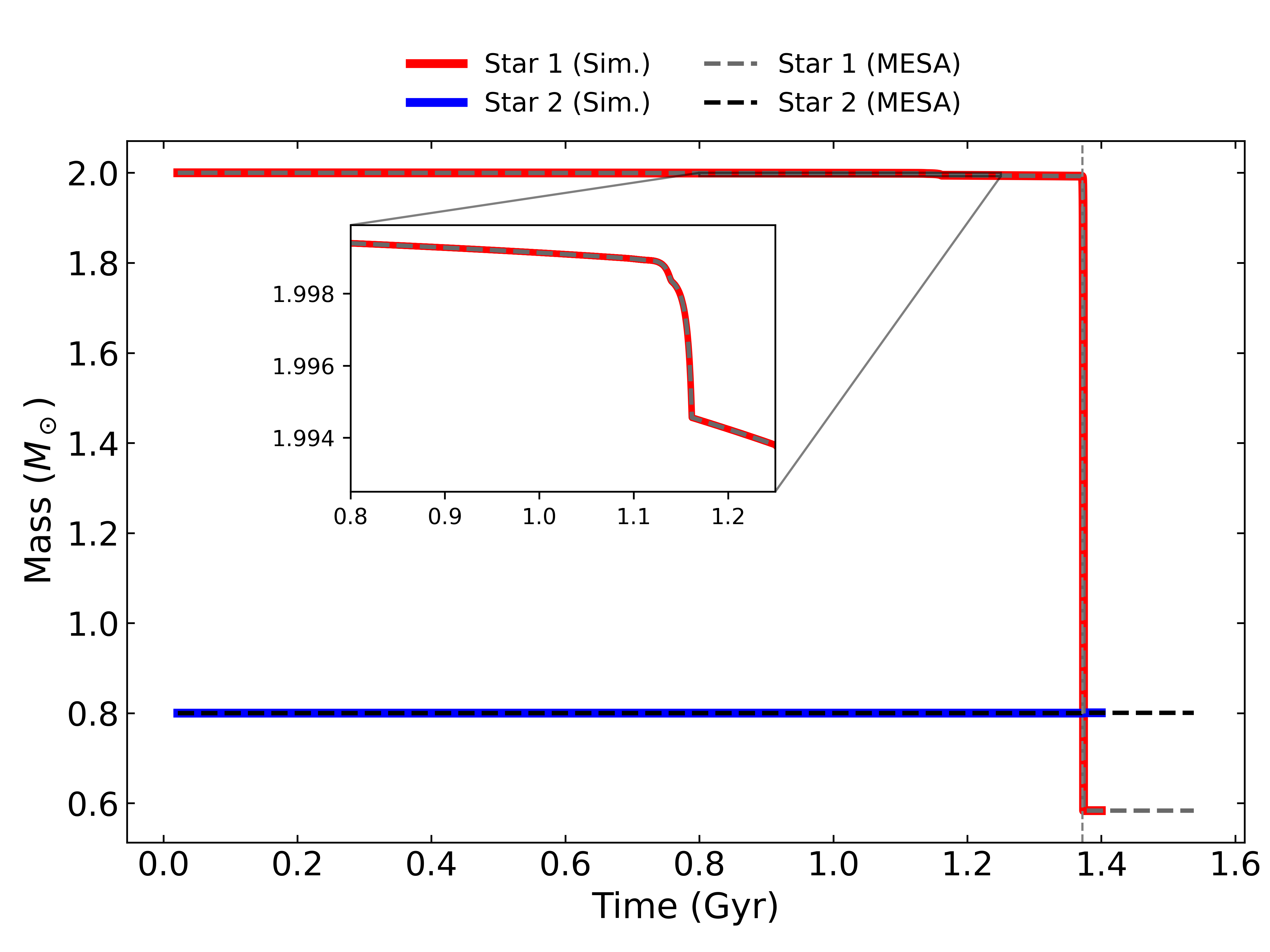}
\caption{
Time evolution of stellar masses obtained by \texttt{MESA} stellar evolution code (dashed lines) for a binary system described in Section~\ref{Sec:MESAAssumptions}. 
Solid lines represent the \texttt{MESA} data interpolated in $N$-body code. 
The vertical dashed line at $t\approx1.37$~Gyr indicates the moment the mass loss from the primary becomes significant. The zoom shows the loss of mass experienced by the primary star as it evolves. 
}
\label{fig:only_stars}
\end{figure}

Figure~\ref{fig:only_stars} shows the masses of the two stars as a function of time, obtained from \texttt{MESA} simulations. The primary star has an initial mass and radius of $2\,M_{\odot}$ and $1.65\,R_{\odot}$, respectively, and evolves into a white dwarf with a final mass and radius of $\sim0.58\,M_{\odot}$ and $\sim0.013\,R_{\odot}$.  
The secondary star remains in the main sequence with a mass of $0.8\,M_{\odot}$ and a radius of $\sim0.7\,R_{\odot}$. \added{The initial semi-major axis of the binary system is 100~au, and this separation increases to 202 au.}
The figure also shows the results of the stellar mass and radius interpolation over the entire stellar evolution process, incorporating these time-dependent quantities into the \texttt{REBOUND} code (solid lines in blue and red).

As highlighted in the close-up of the figure, the primary star undergoes continuous mass loss throughout its evolutionary track, steadily losing mass over the full 1.4\,Gyr interval covered by the simulations. 
However, it is during the thermally pulsing AGB phase that the primary loses its envelope in strong episodes of mass loss due to instabilities in the helium-burning layer around the carbon and oxygen core. Afterward, the primary loses a significant amount of mass, shedding its entire envelope, then evolves into a post-AGB star (see Figure~\ref{fig:HRdiagram}). In this phase, the star's temperature rises high enough, and the ejected material may be observed as a planetary nebulae (i.e., ionized gas surrounding a white dwarf). Once energy generation stops, the object becomes a white dwarf and enters the cooling sequence (i.e., it simply cools down as there are no more nuclear reactions to generate energy). 
The winds and this planetary nebulae originating from the evolution of the primary are not considered here. The wind of the secondary   star has also a negligible influence on the chemical changes of the atmosphere of the white dwarf due to the large separation between the stars.

\subsection{Simulation Setup: Evolving Binary Stars and Jupiter planets}

In this section, we perform a simulation using \texttt{REBOUND} that includes a Jupiter-like planet and two stars, which evolve following the stellar evolution shown in Figure~\ref{fig:only_stars}.
The planet distance is initially set as 5~au  from the primary star and its eccentricity is set as zero.

\begin{figure}
\includegraphics[width=\columnwidth]{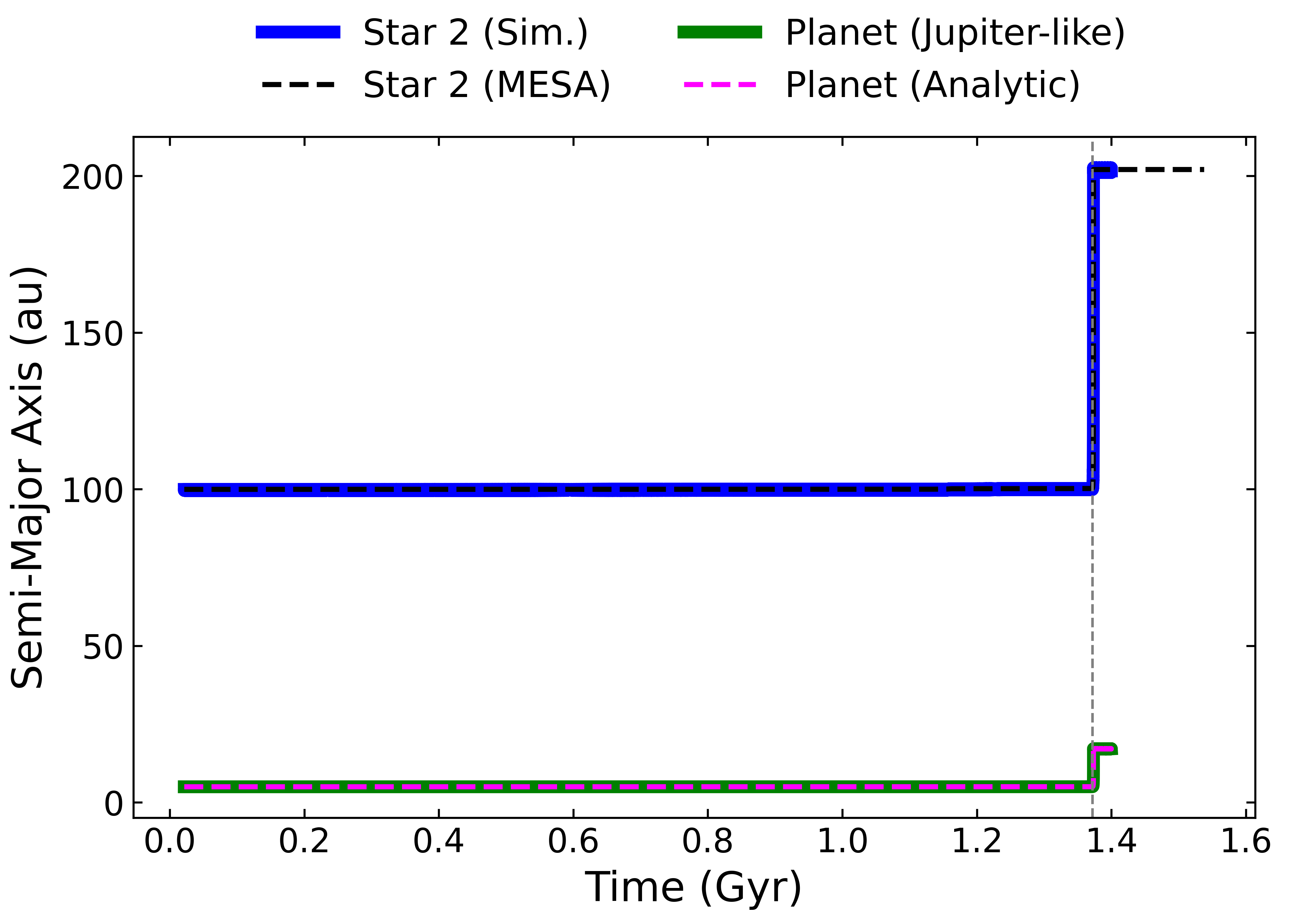}  \includegraphics[width=\columnwidth]{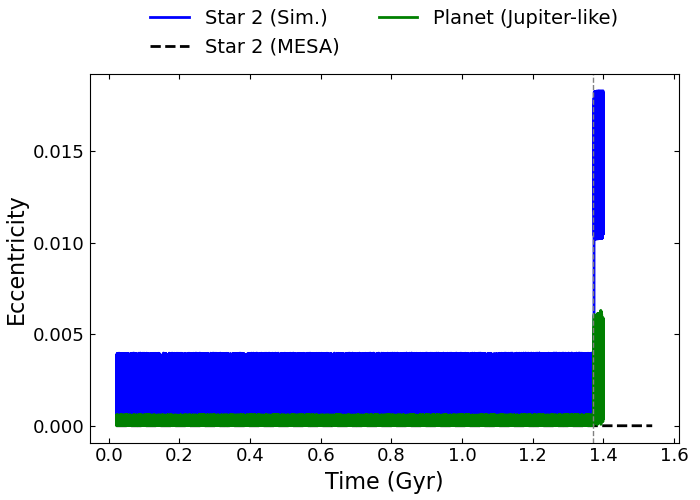}
\caption{
Time evolution of the  orbital elements in a binary system hosting a Jupiter-like planet initially placed at 5~au. 
The blue lines correspond to the $N$-body simulation data, the black dashed line represents the \texttt{MESA}-predicted orbit of the secondary star. The inclination and eccentricity of the secondary are initially set to zero.
\textbf{Top:} Semi-major axis evolution (in au) for the secondary star and the Jupiter-like planet. The analytical equation for the planet’s orbit due to the  mass loss of the primary star is shown in magenta dashed line.
\textbf{Bottom:} Orbital eccentricities obtained for the same bodies. 
The vertical dashed line at $t\approx1.37$~Gyr indicates the onset of significant mass loss of the primary star.
}
\label{fig:example_simulation_axt}
\end{figure}

Figure~\ref{fig:example_simulation_axt} shows the time evolution of the orbital elements of the secondary star and the Jupiter-like planet. After stellar evolution begins, the planet undergoes dynamical changes driven by the host star’s rapid mass loss that it evolves into a post-AGB star. The planet survives all evolutionary phases of the host star and remains in a stable orbit. The semi-major axis and eccentricity of the planet reach approximately 13~au and 0.007, respectively.

The evolution of the semi-major axis of the secondary star obtained with \texttt{REBOUND} (blue solid line in Figure~\ref{fig:example_simulation_axt}) is consistent with the evolution of the semi-major axis predicted by \texttt{MESA} (black dashed line), even when a Jupiter-like planet is included in the system. The eccentricity evolution of \texttt{MESA} does not self-consistently integrate the system as a $N$-body code; consequently, the stellar eccentricity computed by \texttt{MESA} does not include perturbations from the planet.

The expected evolution of the semi-major axis of the planet by inputting the binary masses is given analytically by Equation~\ref{semimajor_planet_eq} \citep[see,][]{Xing_2025}. 

\begin{equation}
a_{f} = a_{i} \frac{M_{1,i} + M_{2,i}}{M_{1,f} + M_{2,f}},
\label{semimajor_planet_eq}
\end{equation}
\noindent where $a_{f}$ and $a_{i}$ represent the orbital separations between the planet and the binary after and before mass transfer, respectively. $M_{1,i}$, $M_{1,f}$ are the masses of the primary and $M_{2,i}$ and $M_{2,f}$ are the masses of the secondary before and after mass transfer, respectively. 

Figure~\ref{fig:example_simulation_axt} also shows (magenta dashed line in the top panel) that the analytical calculation aligns with the evolution of the  planet's semi-major axis obtained from the $N$-body simulation due to mass loss of the primary star. The mass loss of the primary star is assumed to have no effect on the mass of the planet, but drives the orbital response of the system.

\subsection{Simulation Setup: Jupiter, Saturn, and other planets}

We aim to investigate the dynamical evolution of different planetary systems composed of giant planets$-$one Jupiter-like and one Saturn-like planet$-$as well as lower-mass planets, including super-Earths and ice giants (analogous to Uranus and Neptune), under the gravitational influence of the evolving binary stars. The adopted planetary architectures are intentionally inspired by the giant-planet configuration of the Solar System and are designed to serve as analogs for its formation and early dynamical evolution \citep{Tsiganis2005, Nesvorny2012,Izidoro2015}. 

To construct a diverse set of planetary systems, we designed a suite of nine simulation setups by varying the initial total mass of the solid disk (30, 45, 50, and 60~$M_{\oplus}$) and the number of planets per system (\added{1}, \added{2}, \added{4}, \added{9}, \added{14} and \added{19}), \added{which individual planets masses are 15~$M_{\oplus}$, 6~$M_{\oplus}$, 10~$M_{\oplus}$, 12~$M_{\oplus}$, 5~$M_{\oplus}$, 6~$M_{\oplus}$, 4~$M_{\oplus}$, 3~$M_{\oplus}$.} These setups aim to explore formation pathways that could yield final planet masses comparable to those of Uranus and Neptune in our Solar System \citep{Izidoro2015,2023Guga}. 
Table~\ref{tab:simulation_mass} summarizes \added{these} nine simulation configurations considered in this study, defined by variations in the initial number of planets, the total masses of the solid disk, and the resulting individual planet masses. Each configuration is represented by a simulation tag that encodes the initial number of planets and the total disk mass in Earth masses. For example, Sim.~TAG~=~${\added{1}-30}$ denotes a \added{S-type orbit} containing one Jupiter-like planet, one Saturn-like planet, and \added{one} additional planet with individual mass of $15\,M_{\oplus}$, yielding a total disk mass of $30\,M_{\oplus}$.

For each configuration, we included a pair of Jupiter-like and Saturn-like planets in a compact configuration, initially placed on circular orbits at 5.0~au and 6.55~au, respectively, in a 3:2 mean-motion resonance (MMR). The exoplanets were initially positioned using the following procedure: we first calculated Saturn-like's Hill radius and added to it an exoplanet in Hill radius calculated at \added{Saturn-like's} heliocentric distance multiplied by a random factor between 5 and 10. The following exoplanets were positioned following the same procedure, first calculating the Hill radius of the planet at its random position and adding the Hill radius multiplied by another random factor between 5 and 10.
This is important because the formation of the planetary embryos by pebble accretion typically creates a gravitational stable separation between their neighbors by a factor of 5 Hill radius \citep{Kokubo2000}.\added{Therefore, the initial semi-major axis of the most distant exoplanets can range from 9 to 42~au, depending on the number of initial planets with masses similar to those of Uranus and Neptune.} 

We consider that, in our simulation, the stars have evolved and are no longer embedded in a protoplanetary gas disk or a massive planetesimal disk. Jupiter-like and Saturn-like planets are placed on orbits similar to those observed in the Solar System, and a population of super-Earth and Neptune-mass planets are included (see Table~\ref{tab:simulation_mass}). For comparison, we also performed control sample counterparts with no stellar evolution, that is, the stellar parameters, such as radii and masses, are held constant.

\begin{deluxetable}{cccc}
\digitalasset
\tablewidth{0pt}
\tablecaption{Initial parameters for the simulated planetary systems, including the number of planets, total solid disk mass, and individual planet masses. Each system also includes a Jupiter-like and Saturn-like planets initially placed on circular orbits at 5.0~au and 6.55~au, respectively. \label{tab:simulation_mass}}
\tablehead{
\colhead{Sim.} & \colhead{Initial} & \colhead{Total Disk} & \colhead{ Individual planet}\\
\colhead{TAG} & \colhead{N. of planets} & \colhead{Mass ($M_{\oplus}$)} & \colhead{Mass ($M_{\oplus}$)}
} 
\startdata
{\added{1}-30}  & \ralgn{\added{1}}  & 30 & \ralgn{15} \\
{\added{2}-45}  & \ralgn{\added{2}}  & 45 & \ralgn{15} \\
{\added{4}-30}  & \ralgn{\added{4}}  & 30 & \ralgn{6}  \\
{\added{4}-50}  & \ralgn{\added{4}}  & 50 & \ralgn{10} \\
{\added{4}-60}  & \ralgn{\added{4}}  & 60 & \ralgn{12} \\
{\added{9}-50} & \ralgn{\added{9}} & 50 & \ralgn{5} \\
{\added{9}-60} & \ralgn{\added{9}} & 60 & \ralgn{6} \\
{\added{14}-60} & \ralgn{\added{14}} & 60 & \ralgn{4} \\
{\added{19}-60} & \ralgn{\added{19}} & 60 & \ralgn{3} \\
\enddata
\end{deluxetable}

\section{Data analysis and discussion} \label{results}

\subsection{Evolution of the planets} \label{Sec:Results_evolutions}

\begin{figure*}[ht!]  
\gridline{\fig{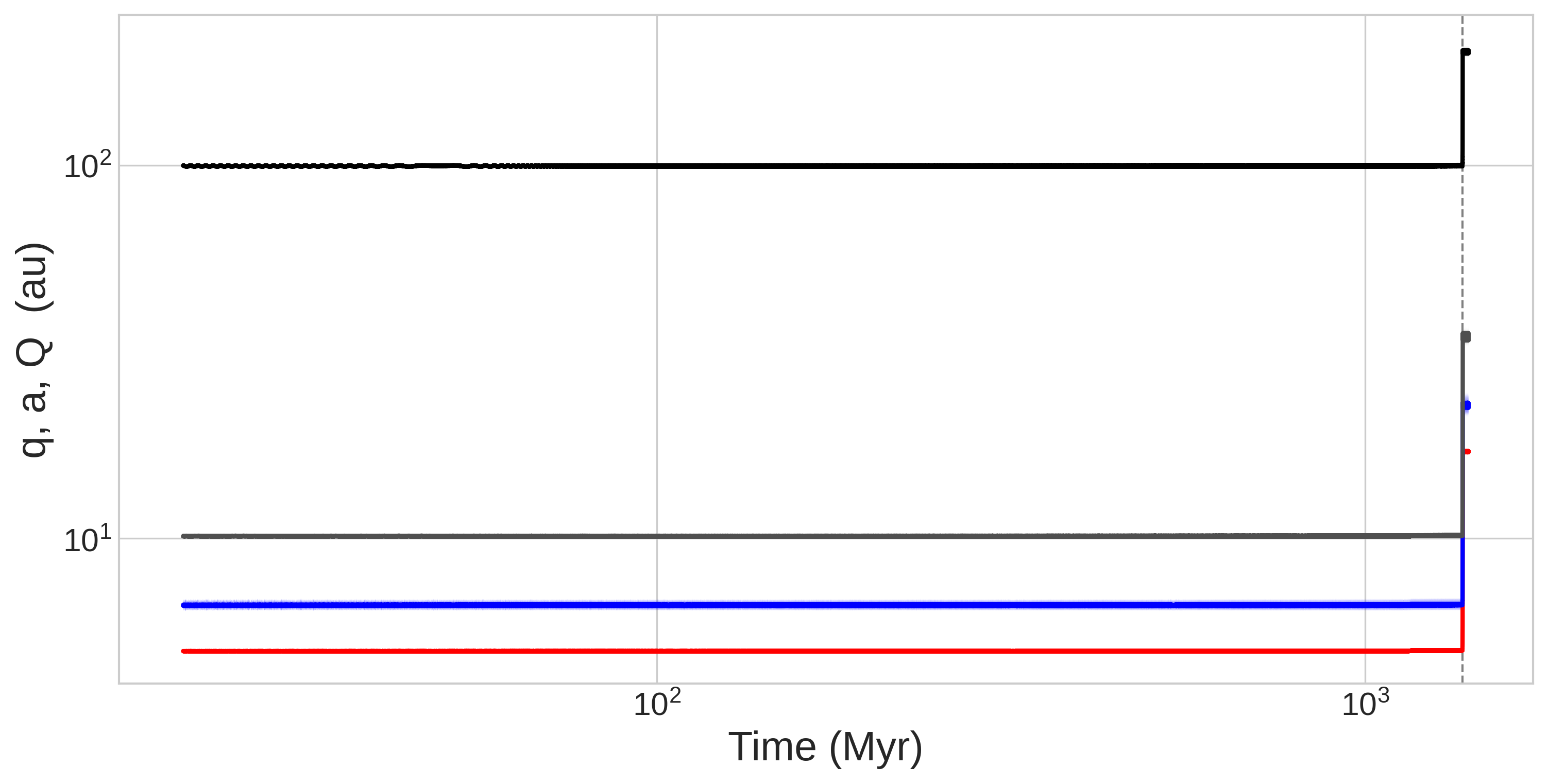}{0.53\textwidth}{(a)}
\fig{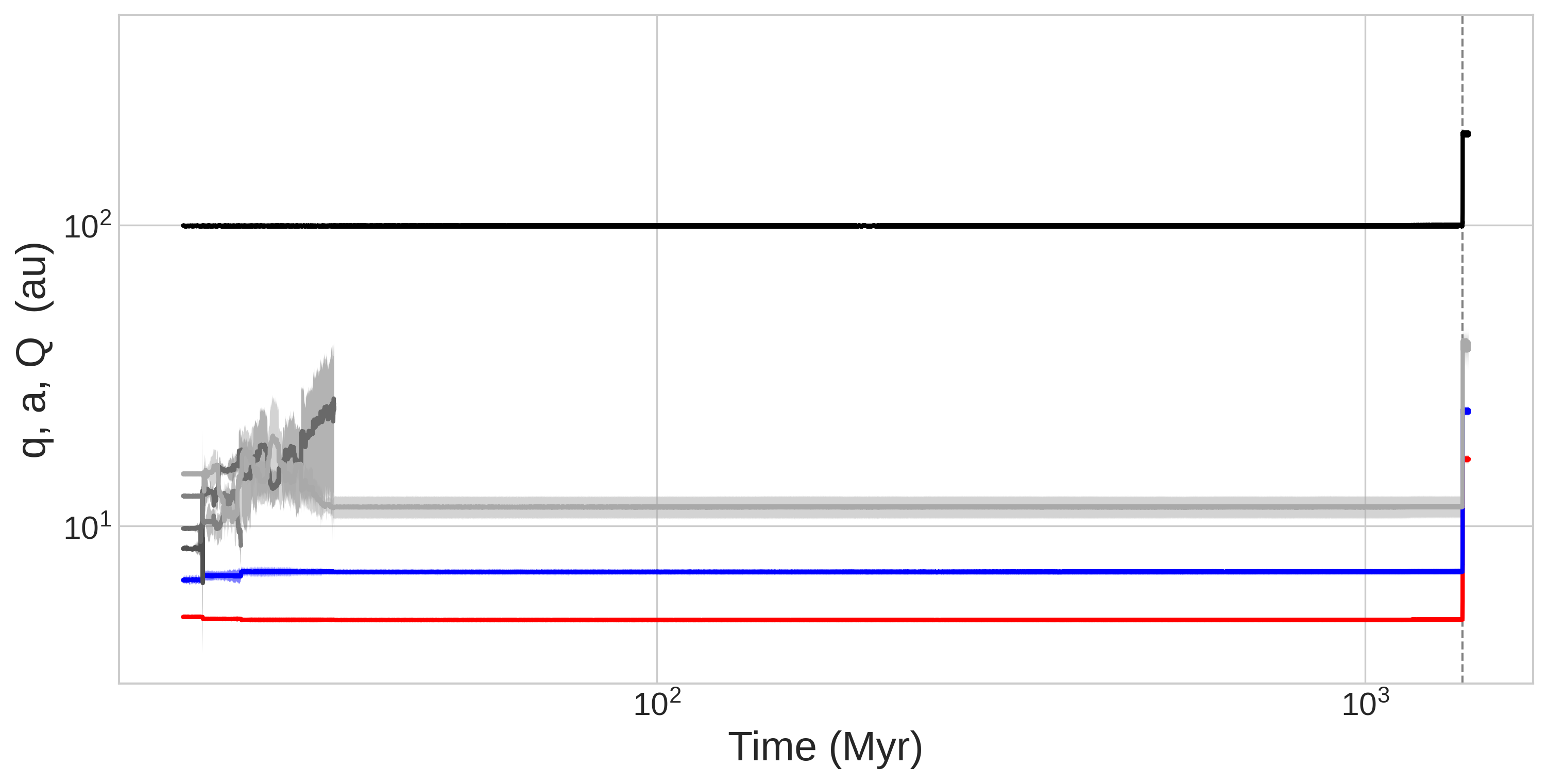}{0.53\textwidth}{(b)}}
\qquad\\
\gridline{\fig{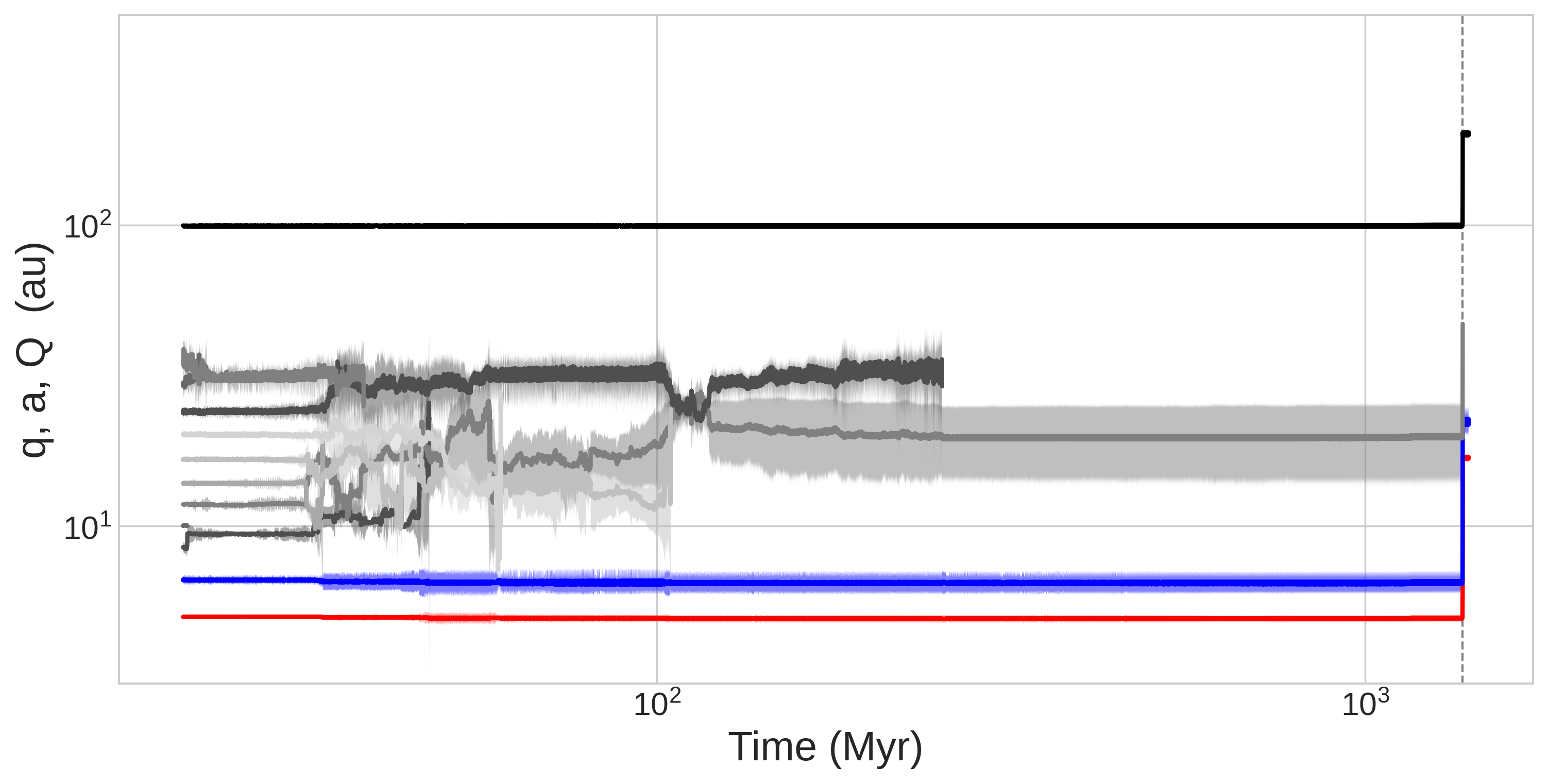}{0.53\textwidth}{(c)}
\fig{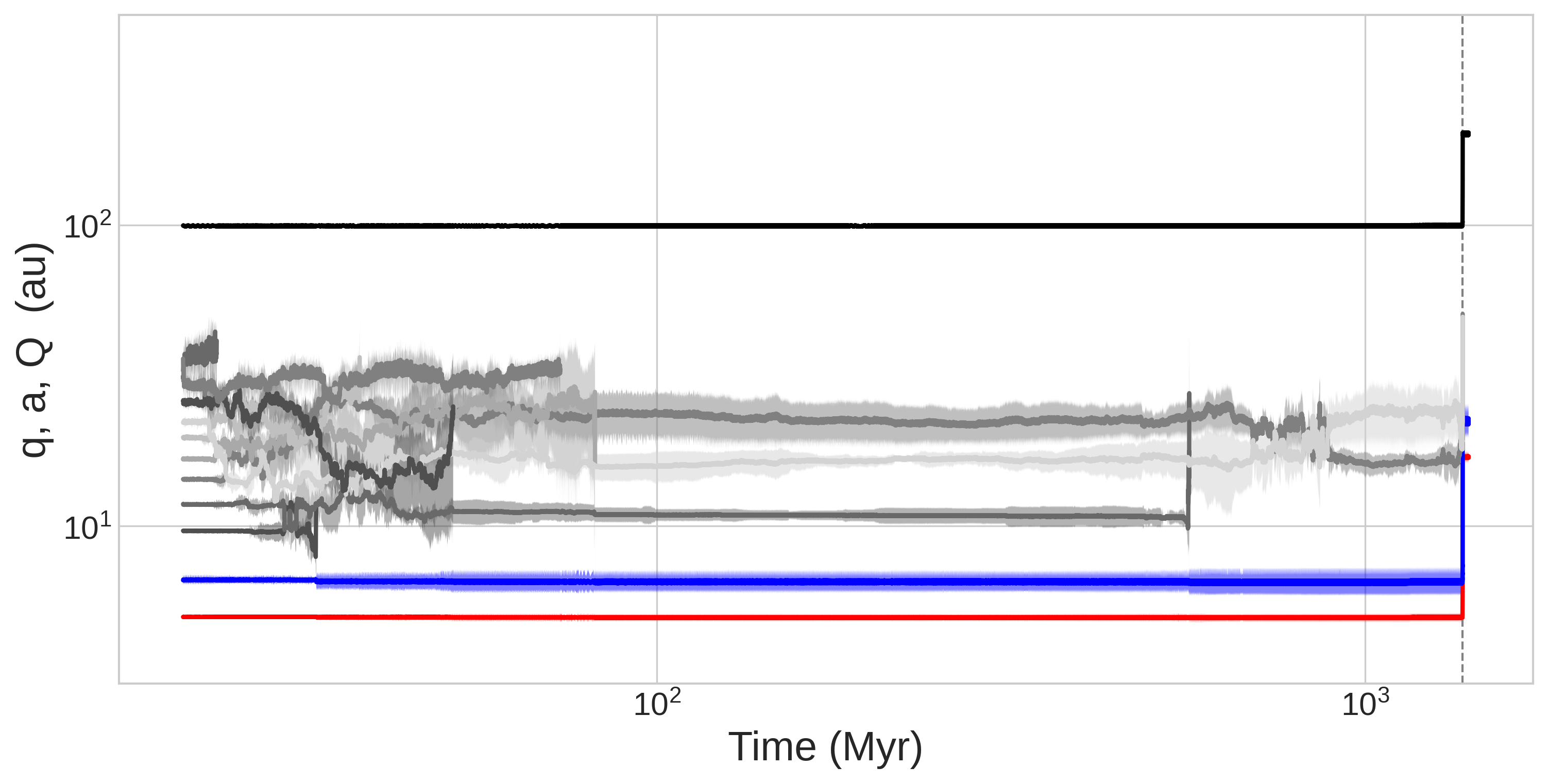}{0.53\textwidth}{(d)}}
\caption{Temporal evolution of the planet’s periapsis (q), semi-major axes (a), and apoapsis (Q) plotted together for each planet in evolved binary simulations. The inclination and eccentricity of the secondary are set to zero.
Panels (a), (b), (c), and (d) show the simulations starting with 1, 4, 14, and 19 exoplanets (gray lines) beyond Jupiter-like (red line) and Saturn-like (blue line) planets. Exoplanets ejected before a 100-timestep are excluded from the simulation and it is not shown here. The vertical dashed line at $t\approx1.37$~Gyr indicates the onset of significant mass loss from the primary star.}
\label{fig:temporal_evolution}
\end{figure*}

\begin{figure*}[ht!]  
\gridline{\fig{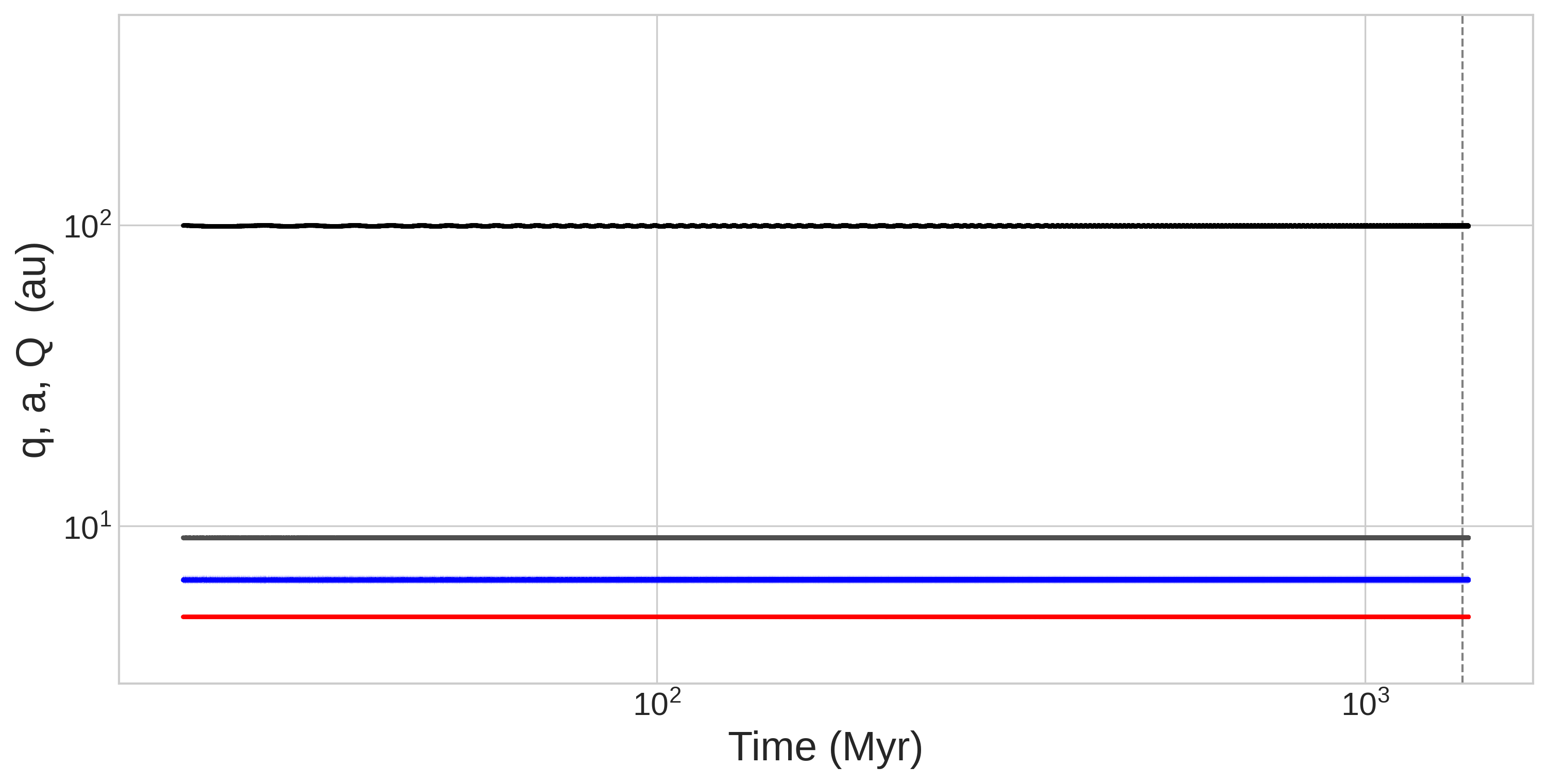}{0.53\textwidth}{(a)}
\fig{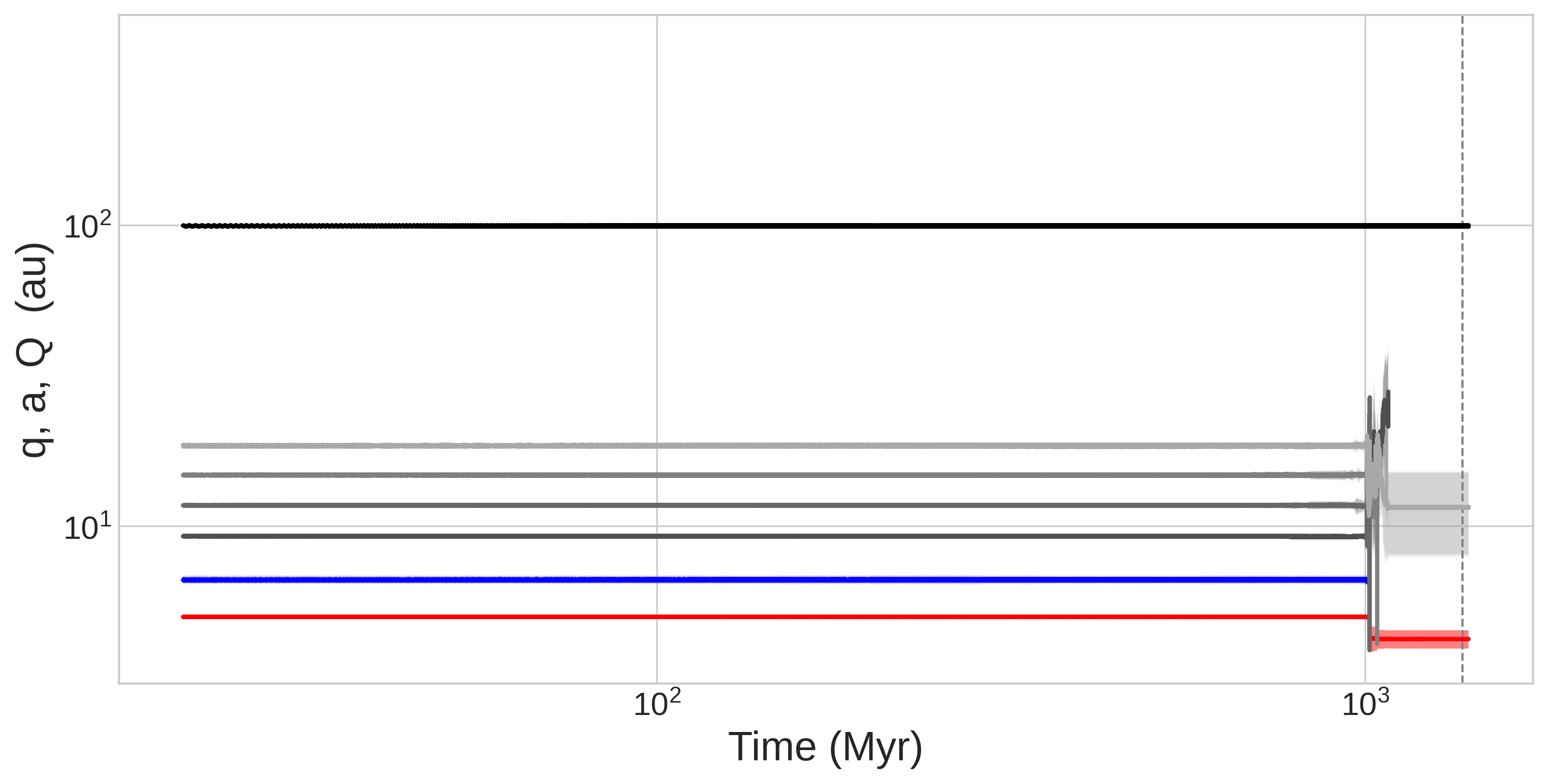}{0.53\textwidth}{(b)}}
\qquad\\
\gridline{\fig{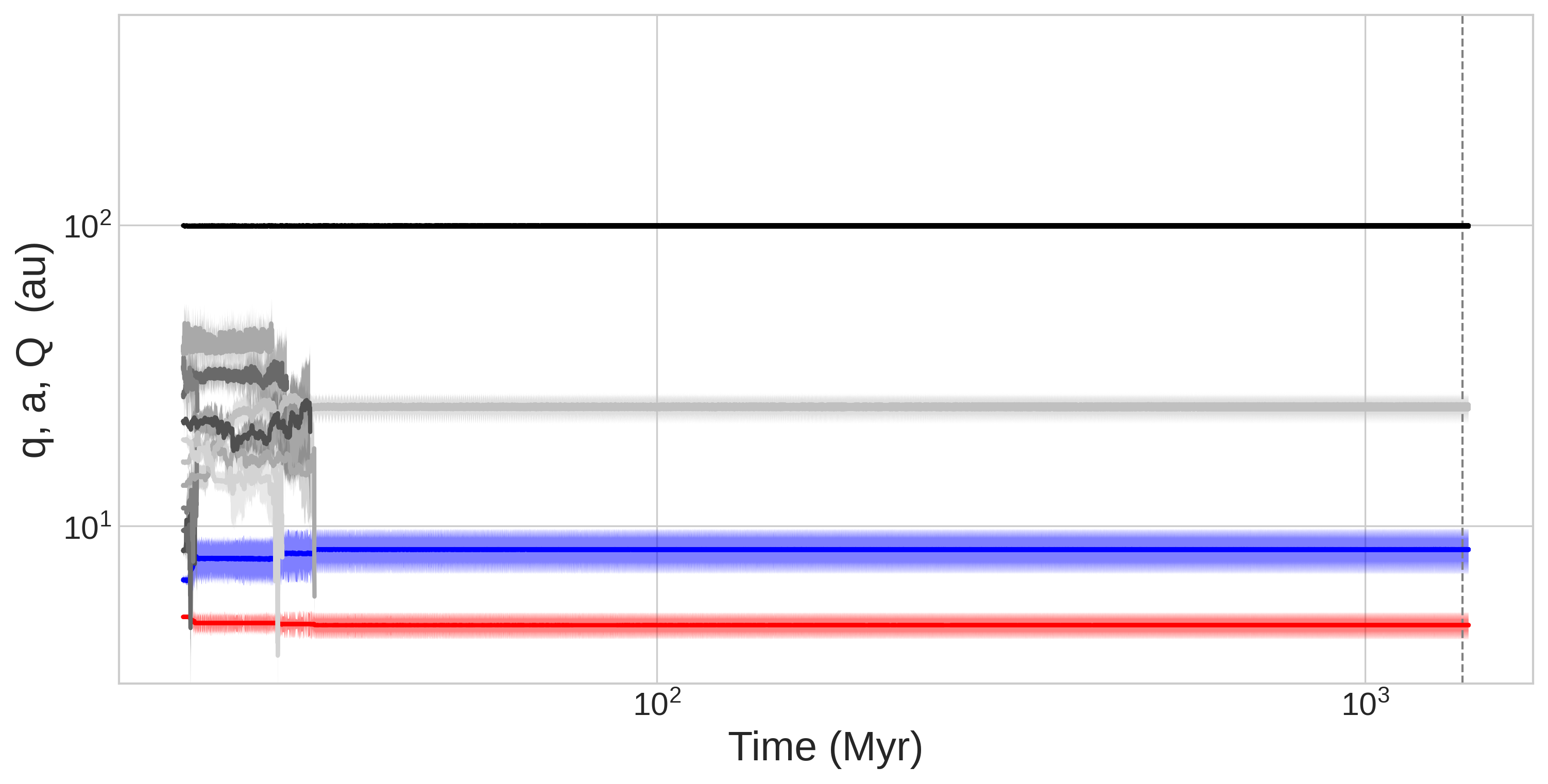}{0.53\textwidth}{(c)}
\fig{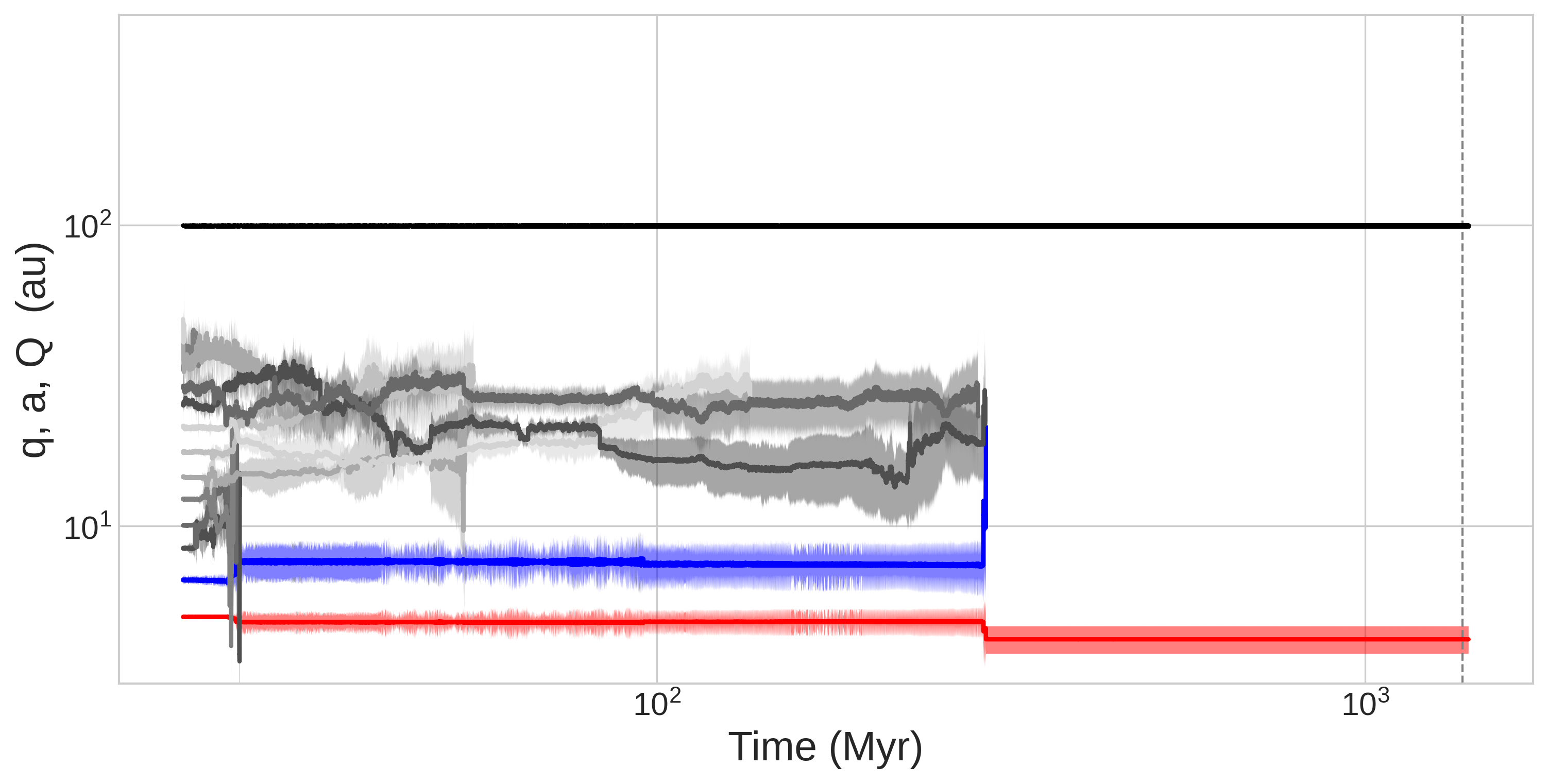}{0.53\textwidth}{(d)}}
\caption{Temporal evolution of the planet’s periapsis (q), semi-major axes (a), and apoapsis (Q) plotted together for each planet in non-evolved binary simulations corresponding to the same simulations in Figure~\ref{fig:temporal_evolution}. The inclination and eccentricity of the secondary are set to zero. Panels (a)-(d) show the simulations starting with 1, 4, 14, and 19 exoplanets (gray lines) beyond Jupiter-like (red line) and Saturn-like (blue line) planets. Exoplanets ejected before a 100-timestep are excluded from the simulation and it is not shown here. The vertical dashed line at $t\approx1.37$~Gyr indicates the onset of significant mass loss from the primary star.}
\label{fig:temporal_nonevolution}
\end{figure*}

Figures~\ref{fig:temporal_evolution} and~\ref{fig:temporal_nonevolution} show the time evolution of the periapsis ($q$), semi-major axis ($a$), and apoapsis ($Q$) of the planets and the secondary star (black line) for four different setups. Panels (a)--(d) show simulations starting with \added{1}, 4, \added{14}, and \added{19} planets beyond Jupiter-like (red line) and Saturn-like (blue line). \added{Some of these exoplanets are ejected before a 100-timestep and are therefore excluded from the simulations, meaning they are not shown here.} The results that include stellar evolution are shown in Figure~\ref{fig:temporal_evolution}, while the corresponding cases without stellar evolution are shown in Figure~\ref{fig:temporal_nonevolution}. It is important to note that the stars lose their mass continuously, producing secular effects as can be seen in the semi-major axes of the exoplanets.

We noted that, in all cases, the highest number of close interactions (e.g., close encounters) among the giant planets and the other planets occurs within the first $\sim$100~Myr, except in simulations starting with a small number of planets, independent of stellar evolution. During this phase, ejections, collisions, and merges among these bodies occur frequently. 

The dynamical evolution of the planetary system can be divided into three main stages. 
The first stage is characterized by the planets interacting frequently and strongly. During this phase, close encounters drive the planetary eccentricities to large values. This initial stage is triggered by the compact initial configuration of the system, which enhances secular perturbations and places the planets on closely spaced orbits. As a consequence of these encounters, the planets are rapidly scattered toward the orbital regions of Jupiter-like and Saturn-like planets. This stage typically lasts up to $\sim$100~Myr and its duration depends on the initial number of planets in the system. Therefore, the first stage is intrinsically unstable and is marked by planetary instabilities, including close encounters, collisions, and the ejection of planets. 

The second stage typically occurs between $\sim$100~Myr and $\sim$1300~Myr, when only a few planets remain. During this phase, some exoplanets are captured in mean-motion resonances (MMRs) with Jupiter-like and Saturn-like planets, reaching equilibrium eccentricities that are usually intermediate ($e\lesssim0.4$), while others become sufficiently separated to avoid strong mutual interactions.
There are a very small number of closest interactions between these bodies in this second stage.
The third stage begins after $\sim$1300~Myr and is associated with the significant mass-loss of the primary star as it evolves into a white dwarf. This phase of stellar evolution induces an abrupt expansion of the stellar orbit and a sudden rearrangement of the planetary system, which is not observed in the control sample without evolution regardless of the number of exoplanets (see Figure~\ref{fig:temporal_nonevolution}). As a consequence of the changing gravitational potential, the remaining planets experience orbital excitation, with noticeable increases in semi-major axes and enhanced eccentricity variations. In some cases, this leads to renewed dynamical activity, including close encounters, scattering events, and the ejection of planets, while a small subset of objects can survive on wider dynamically stable orbits. 

\added{The comparison between Figures~\ref{fig:temporal_evolution} and~\ref{fig:temporal_nonevolution} shows that the onset of instability is not controlled by a single mechanism. In the absence of stellar evolution, mutual planet-planet perturbations can already drive a significant radial spreading of the Jupiter-like and Saturn-like planets, especially when the planets are initially close to mean-motion resonances but not deeply protected by a stable resonant chain. In these cases, the system may remain marginally stable for hundreds of Myr, until the resonant configuration is disrupted and orbit crossing occurs.
When stellar evolution is included, the instability is enhanced because stellar mass loss, especially during the AGB phase, causes the planetary orbits to expand and modifies the period ratios between adjacent planets. This can move the system away from its initial resonant configuration and increase the amplitude of eccentricity oscillations. The resulting architecture becomes more vulnerable to both mutual planetary perturbations and perturbations from the binary companion. Therefore, the enhanced instability found in our simulations should be interpreted as the outcome of a coupled process involving stellar mass loss, binary perturbations, and planet-planet interactions, rather than as the consequence of stellar evolution alone.}

\subsection{Statistical analysis}

Table~\ref{rates} provides a comparative analysis of planetary ejection and merge rates derived from evolved and non-evolved binary simulations.  A consistent trend across all simulation tags (Sim.~TAG) shows that ejection rates are significantly higher in simulations involving binary evolution than in non-evolved counterparts. For example, in Sim.~TAG~=~\added{1}$-$30, which have a Jupiter-like planet, a Saturn-like planet, and \added{one} other \added{planet} with masses of 15~$M_{\oplus}$, the ejection rate increases from 0.12 in the non-evolved scenario to 0.44 when binary evolution is included. As the number of planets in the simulation increases, the difference in ejection rates between evolved and non-evolved binaries decreases, ranging from 0.08 to 0.05 for Sim.~TAG equal to \added{9}$-$50, \added{9}$-$60, \added{14}$-$60, and \added{19}$-$60. From the total number of exoplanets in our simulations (\added{8,400}), \added{1,042} exoplanets (including the population of Jupiter-like and Saturn-like planets) are not ejected in an evolved binary. \added{In the non-evolved binary system, the number of not ejected exoplanets is 2,104 (see Panels (a) and (b) in Figure~\ref{histogram}).} 

Systems that remain stable throughout the entire simulation provide a robust framework for isolating the effects of stellar mass evolution. In simulations with non-evolving binary stars, \added{151} of the planetary systems remain completely stable. In contrast, when stellar mass evolution is included, we find only a single system \added{(Sim.~TAG~=~\added{1}$-$30)} that remains dynamically stable. It seems that stable systems do not remain stable when the stars evolve. 

The merge rates remain consistently low throughout the dataset, generally ranging from 0.01 to 0.02. This suggests that in these binary environments, dynamical instability predominantly leads to the ejection of planets rather than mergers or collisions. In systems with binary evolution, 162 mergers were recorded, compared to 115 in non-evolved binary systems. While this indicates a slightly higher incidence of planet-planet collisions following stellar evolution, the overall probability of merging due to gravitational perturbations remains largely consistent between the two groups.

In 81\% of the mergers, the other planets with masses different from Jupiter-like and Saturn-like can be engulfed by the primary star in evolved binary simulations as well as non-evolved simulations.
Following the collision and merger, a total of 14 exoplanets remain in evolved binary simulations, with masses ranging from 0.025 to 0.33~M$_{\text{Jup}}$, and 9 exoplanets with masses ranging from 0.01 to 0.09~M$_{\text{Jup}}$ in non-evolved simulations. Collisions and mergers were not achieved with Jupiter-like or Saturn-like planets, or with the secondary star.

In evolved binary systems, we find that a fraction of \added{12 surviving} exoplanets can cross the orbits of the Jupiter-like planets; \added{11 Saturn-like planets and 1 planet with mass of 15~$M_{\oplus}$}. In comparison with a non-evolved simulation, a total of \added{21 surviving} exoplanets orbit a semi-major axis smaller than that of a Jupiter-like planet, \added{with 5 Saturn-like planets and other planets with different masses.} 
One exoplanet with a mass of 12~$M_{\oplus}$ has a semi-major axis very close to the secondary star (202~au) with $a$~=~201~au, however it does not remain stable until the end of the simulation with stellar evolution. The same situation occurs in the non-evolved simulation, in which only one exoplanet with a mass of 3~$M_{\oplus}$ remains in an orbit of $a$~=~500~au before ejection.

\begin{deluxetable}{ccccc}
\digitalasset
\tablecaption{The ejection and merge rates obtained from simulations with and without binary system evolution. \label{rates}}
\tablehead{
\colhead{Sim.} & \multicolumn{2}{c}{Evolved binary sim.} & \multicolumn{2}{c}{Non-evolved binary sim.}\\
\colhead{TAG} & \colhead{Ejection rate} & \colhead{Merge rate} & \colhead{Ejection rate} & \colhead{Merge rate}
}
\startdata
{\added{1}-30} & 0.44 & 0.01 & 0.12 & $<$0.01 \\
{\added{2}-45} & 0.56 & $-$  & 0.28 & 0.01 \\
{\added{4}-30} & 0.67 & 0.02 & 0.50 & 0.01 \\
{\added{4}-50} & 0.67 & 0.02 & 0.51 & 0.01 \\
{\added{4}-60} & 0.67 & 0.01 & 0.50 & 0.01 \\
{\added{9}-50} & 0.81 & 0.02 & 0.73 & 0.02 \\
{\added{9}-60} & 0.81 & 0.01 & 0.73 & 0.01 \\
{\added{14}-60} & 0.85 & 0.02 & 0.80 & 0.01 \\
{\added{19}-60} & 0.88 & 0.01 & 0.83 & 0.01
\enddata
\end{deluxetable}

\subsubsection{Cumulative distribution of the semi-major axis} \label{CDF_semi_major}

We analyze the cumulative probability distributions of the final orbital elements of the surviving planets in all simulated systems. We separately examine the semi-major axis, eccentricity, and inclination distributions for Jupiter-like planets, Saturn-like planets, and other planets. For each case, we also compare simulations that include stellar evolution with control simulations in which stellar evolution is neglected.

\begin{figure*}[h]
     \includegraphics[width=\columnwidth]{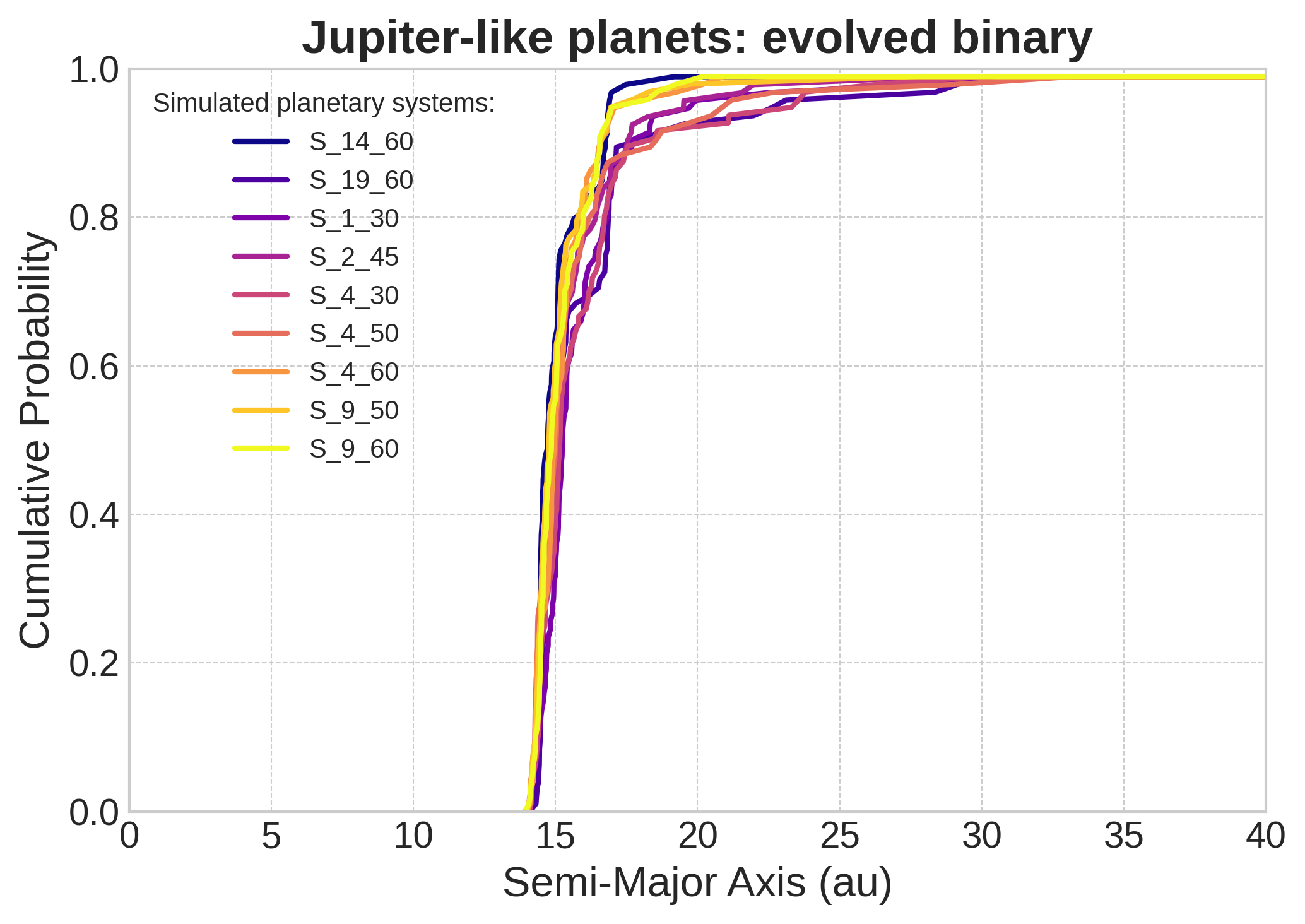}
     \includegraphics[width=\columnwidth]{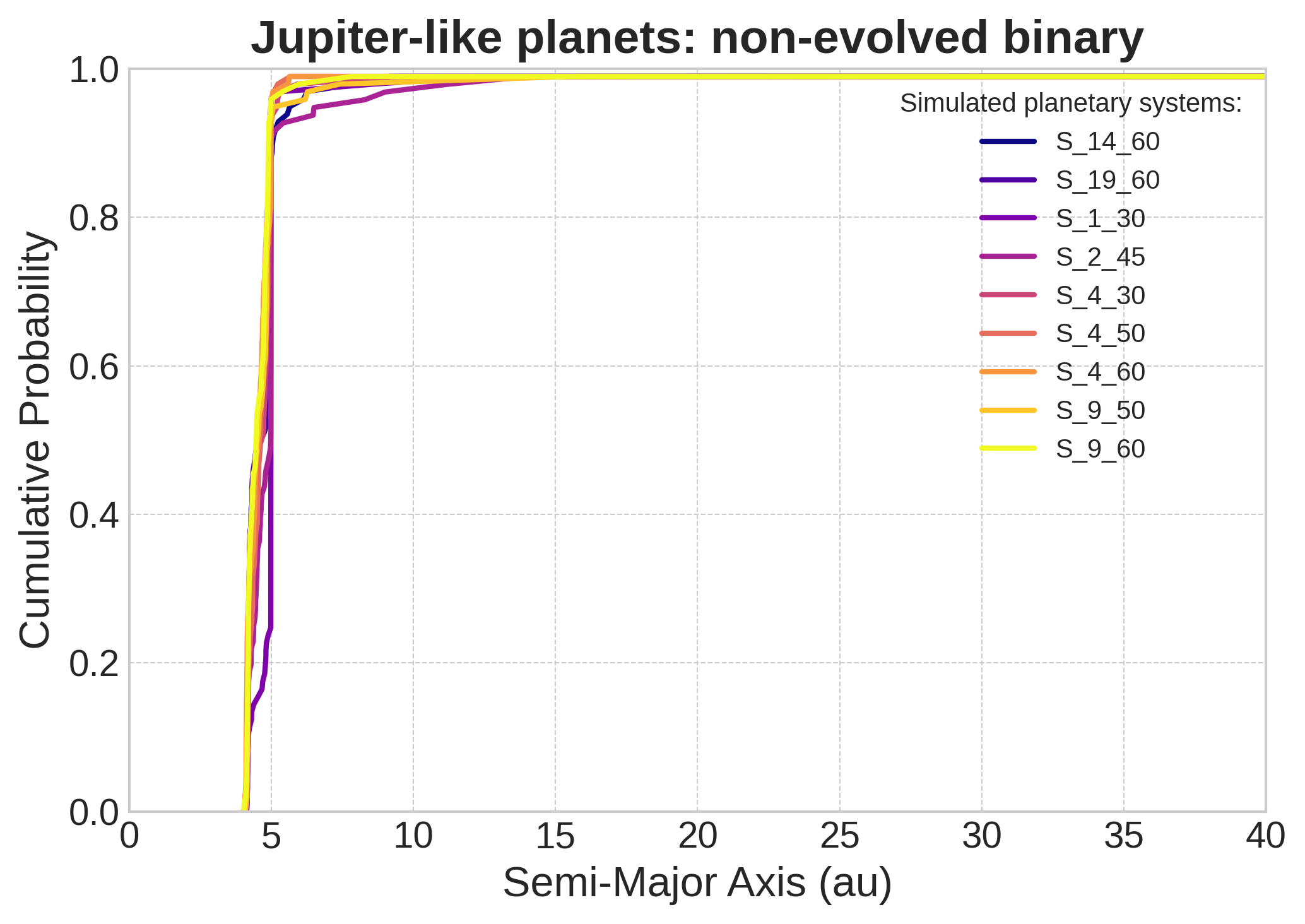}
     \includegraphics[width=\columnwidth]{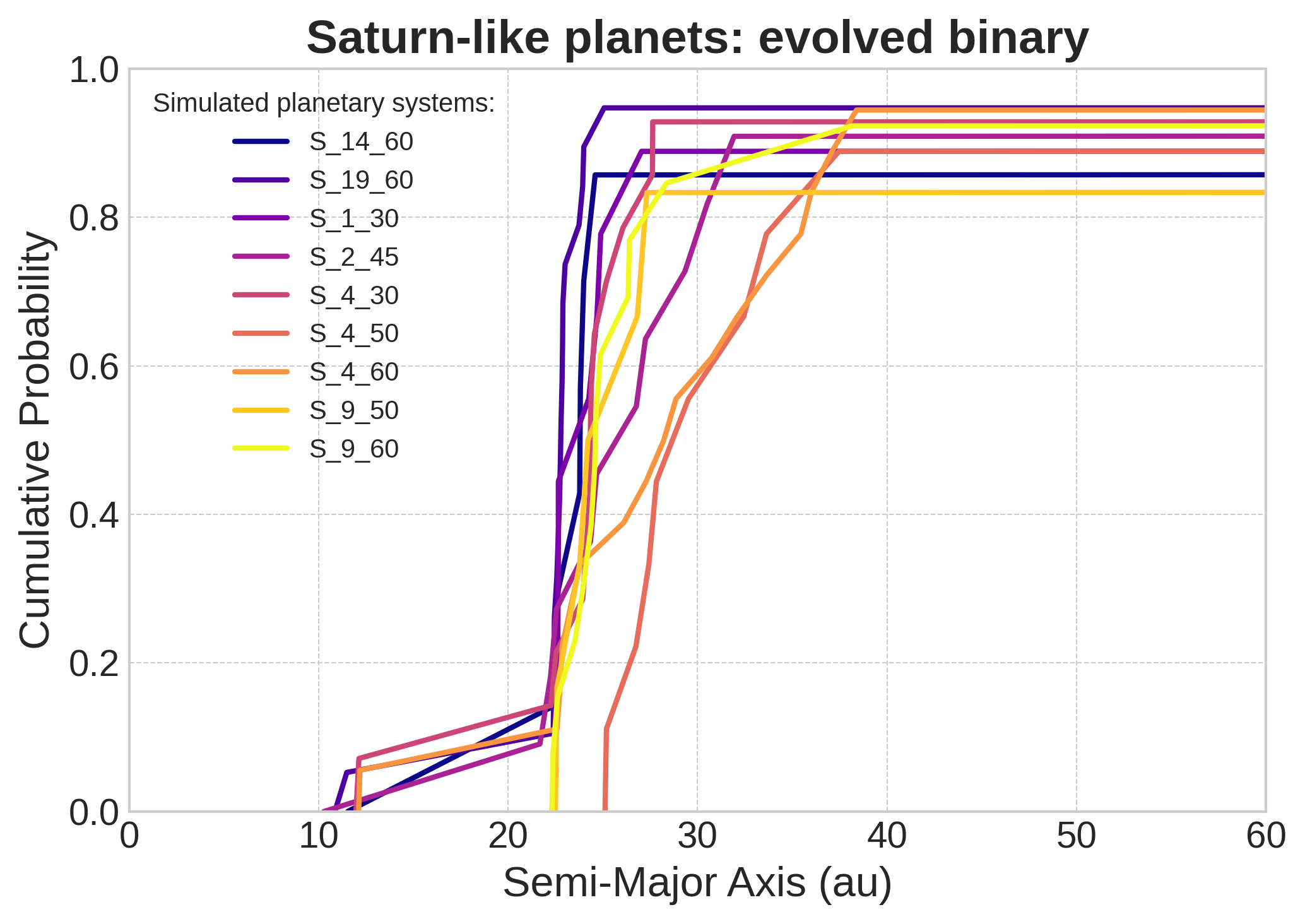}
     \includegraphics[width=\columnwidth]{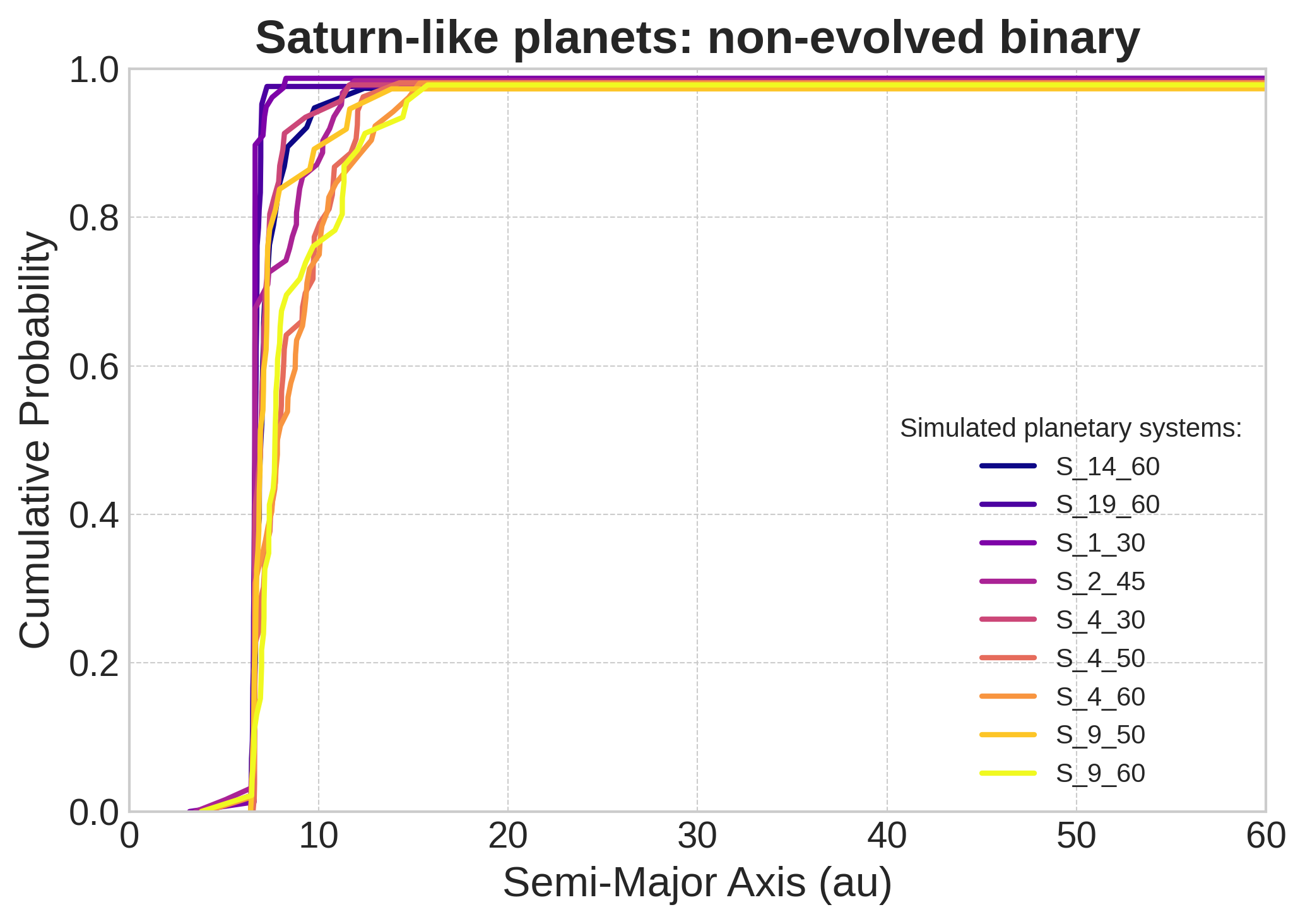}
     \includegraphics[width=\columnwidth]{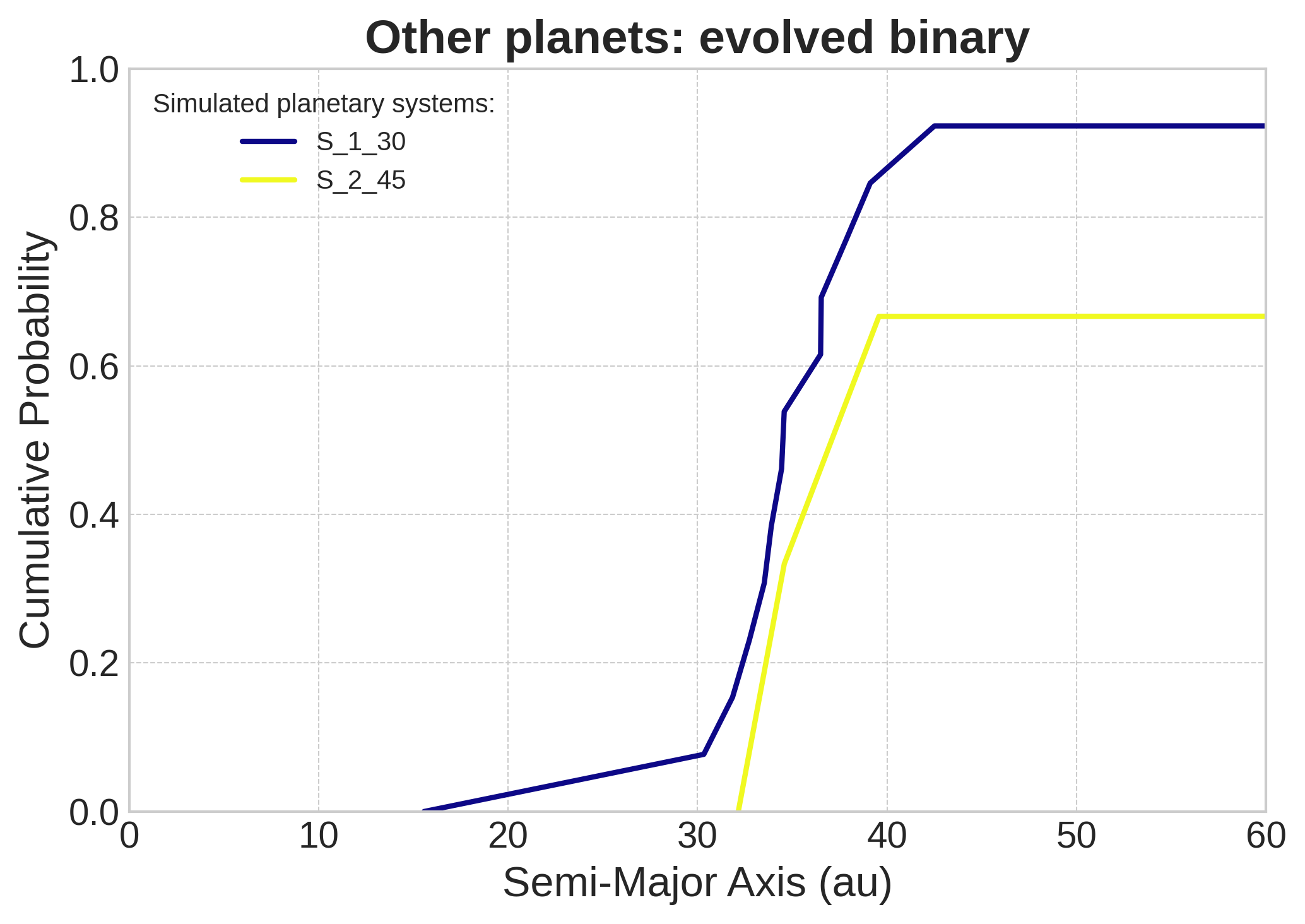}
     \includegraphics[width=\columnwidth]{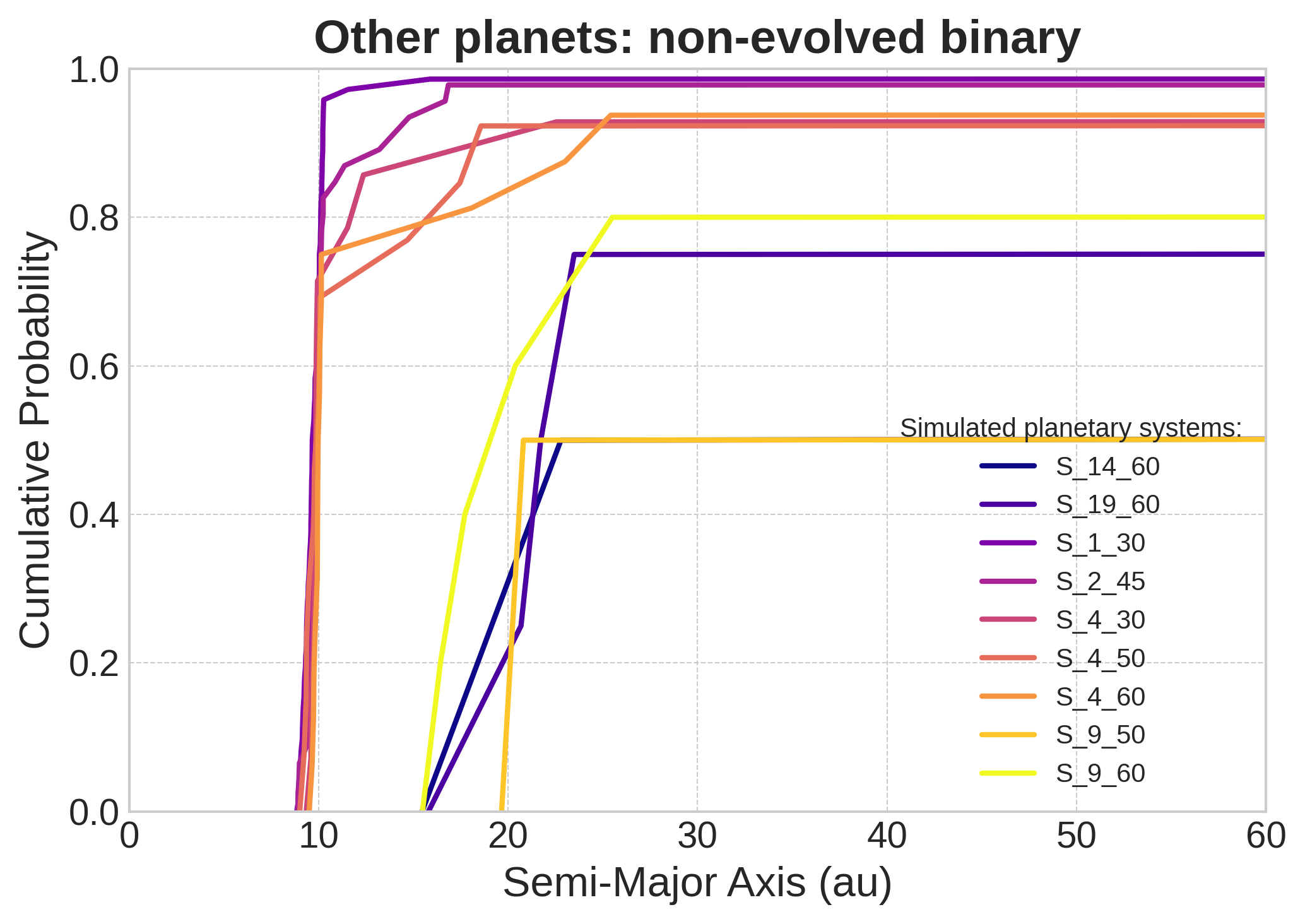}
\caption{
Cumulative normalized distributions of the semi-major axis for surviving Jupiter-like, Saturn-like, and other planets simulated for different planetary systems (see Table~\ref{tab:simulation_mass}). The left panels show the evolved binary simulations and the right panels show the control simulations (non-evolved binary).}
\label{fig:cdf_semimajor}
\end{figure*}

Figure~\ref{fig:cdf_semimajor} shows the cumulative probability distributions of the final semi-major axes of the surviving exoplanets. The left panels correspond to simulations that include stellar evolution, whereas the right panels present the control simulations without evolution.

In the absence of stellar evolution, the final semi-major axis distribution remains strongly concentrated near the initial orbital location of Jupiter-like and Saturn-like planets. Nearly all surviving Jupiter-mass planets remain confined within $\sim$5~--~6~au, and the cumulative probability rises steeply at these distances, rapidly approaching unity. This behavior indicates that, in dynamically stable systems, late-stage planet--planet interactions alone are inefficient at significantly redistributing Jupiter-mass planets.
When stellar evolution is included, the distribution becomes markedly broader. In this scenario, surviving Jupiter-like planets populate a wide range of semi-major axes, extending from $\sim$15~au to beyond 30~au. The more gradual rise of the cumulative probability reflects the combined effects of stellar mass-loss and long-term planet--planet scattering, which together promote orbital expansion and enhanced radial diffusion. Thus, stellar evolution enables a significant restructuring of the orbital architecture of Jupiter-like planets that would otherwise remain compact.

In Saturn-like planets, the cumulative probability distributions show that the control simulations exhibit exoplanets with orbital distances within a range of approximately 7~au up to $\sim$15~au. The cumulative probability increases suddenly with the semi-major axis, indicating a non-redistribution driven by long-term planet--planet interactions. In the evolved simulations, clear differences emerge between configurations with different numbers of planets, particularly regarding the orbital distribution of Saturn-like planets. In simulations with a larger number of planets (e.g., 10 and 20 bodies), Saturn-like survivors typically remain at semi-major axes of about ${20-25}$~au. In contrast, systems initialized with fewer planets tend to produce Saturn-like planets on significantly wider orbits, with semi-major axes reaching ${\sim30-40}$~au. 
These results suggest that Saturn-like planets are systematically displaced to semi-major axes beyond $\sim$10~au, representing a significant difference relative to simulations without binary evolution. This behavior highlights the strong sensitivity of Saturn-like planets to both the initial planetary multiplicity and the dynamical perturbations induced by an evolving binary system.

The cumulative probability distributions of the semi-major axes of surviving planets other than Jupiter-like and Saturn-like (see Figure~\ref{fig:cdf_semimajor}) show that the most surviving planets remain relatively confined in control simulations, with the majority residing within $\sim 15$~au and 25~au for simulations with fewer and more exoplanets, respectively. The steep initial rise in cumulative probability reflects the tendency of these planets to remain in moderately compact orbits despite early dynamical instabilities.
In stellar evolution simulations, the distributions become significantly broader, but only simulations with 2 and 3 exoplanets survived with masses of 15~$M_{\oplus}$. The cumulative probability increases more gradually with the semi-major axis, and a substantial fraction of the planets is found beyond ${30-40}$~au, extending up to $\sim 50$~au. This behavior highlights the strong sensitivity of lower-mass planets to stellar mass-loss, which in combination with planet--planet scattering efficiently reshapes their orbital architecture.

\subsubsection{Cumulative distribution of  the eccentricity} \label{CDF_eccetricity}

\begin{figure*}[ht!]
     \includegraphics[width=\columnwidth]{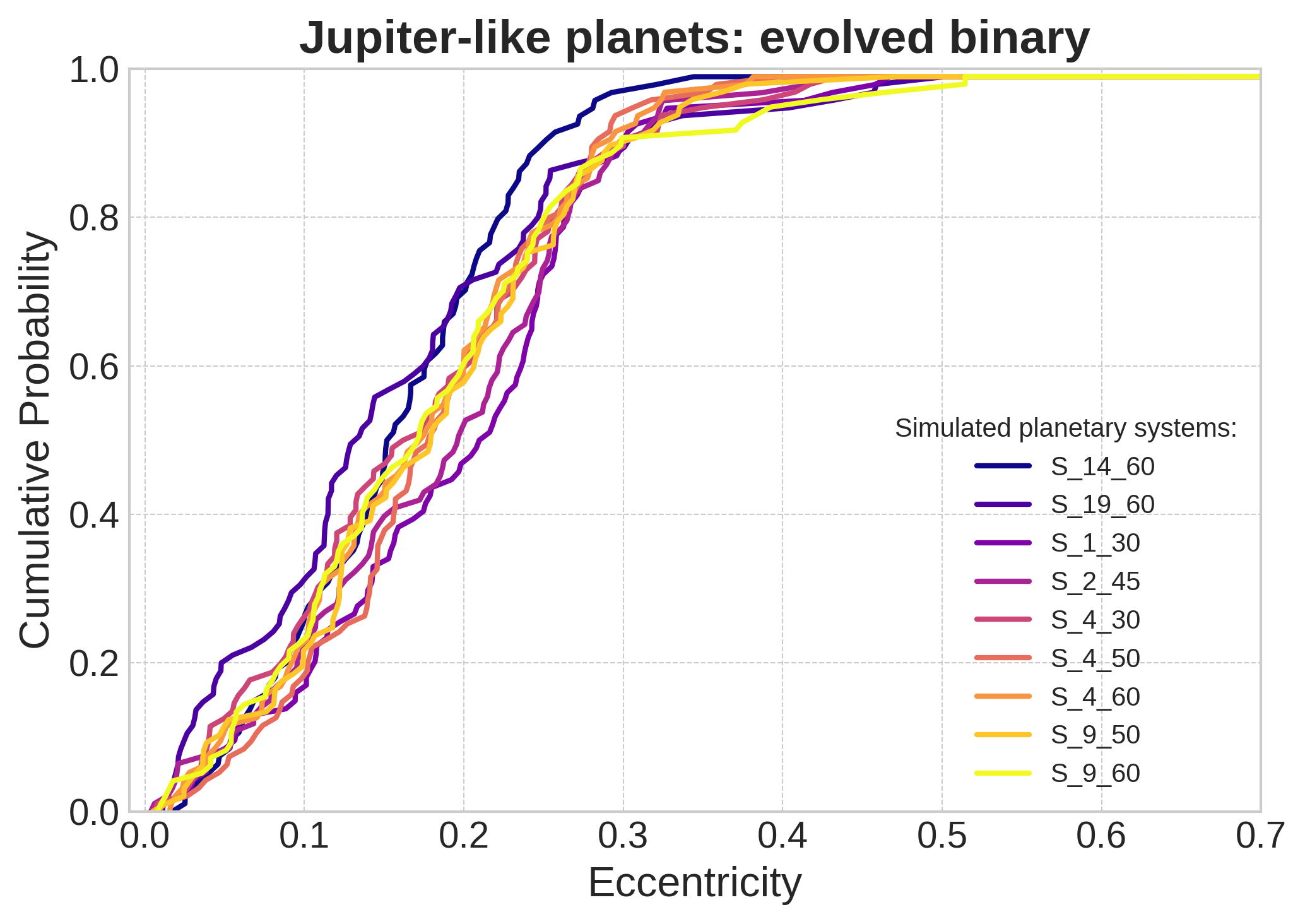}
     \includegraphics[width=\columnwidth]{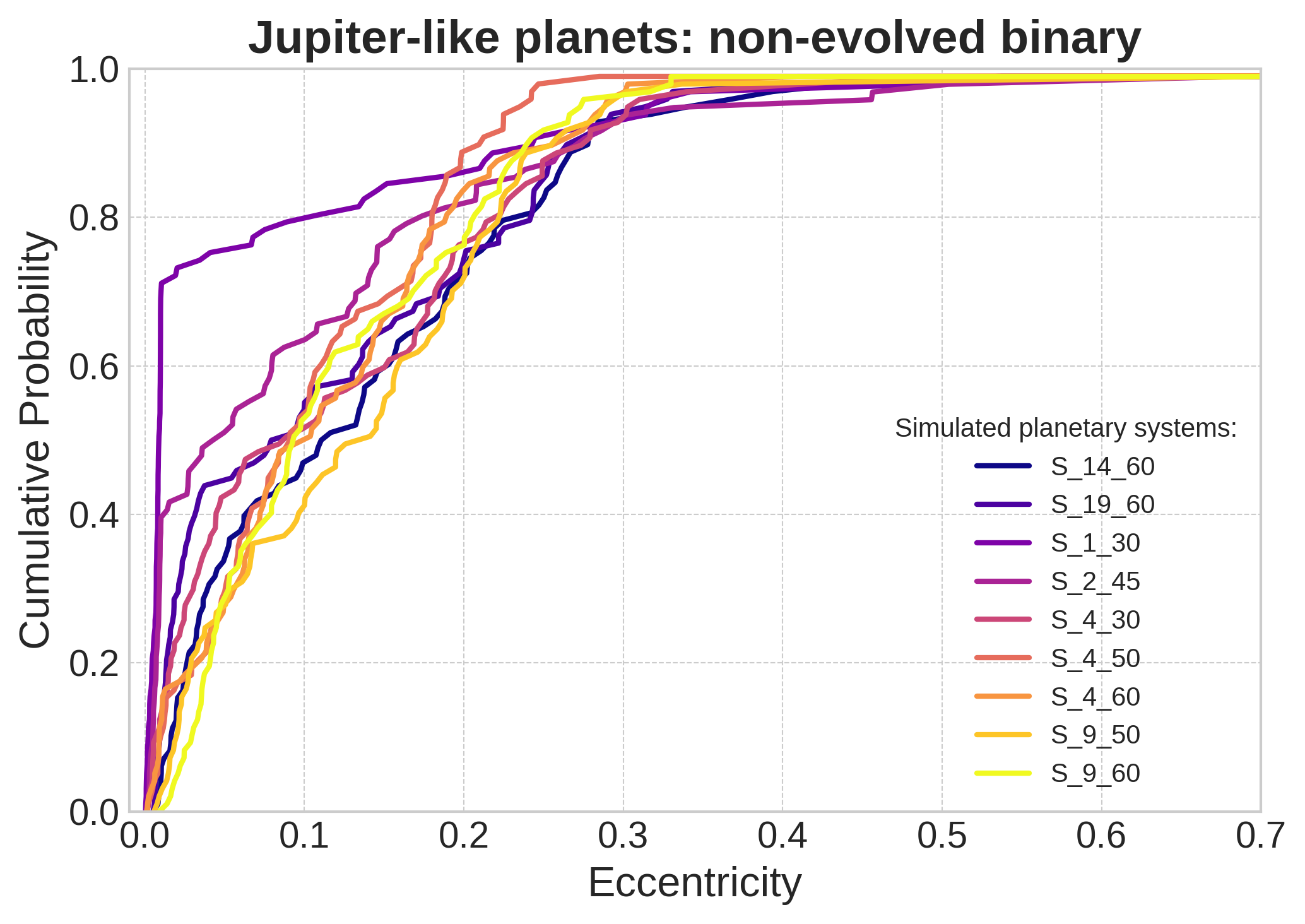}
     \includegraphics[width=\columnwidth]{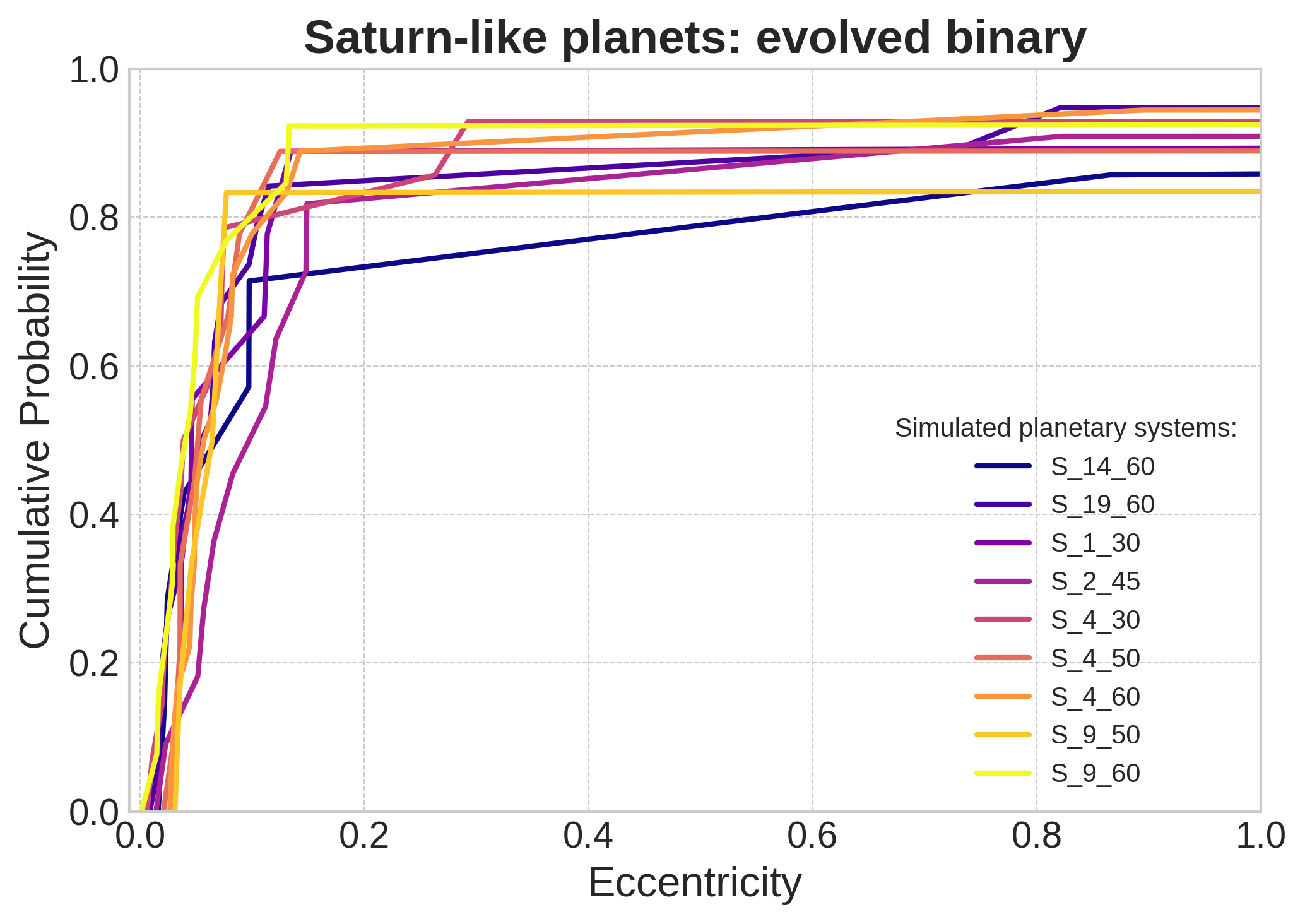}
     \includegraphics[width=\columnwidth]{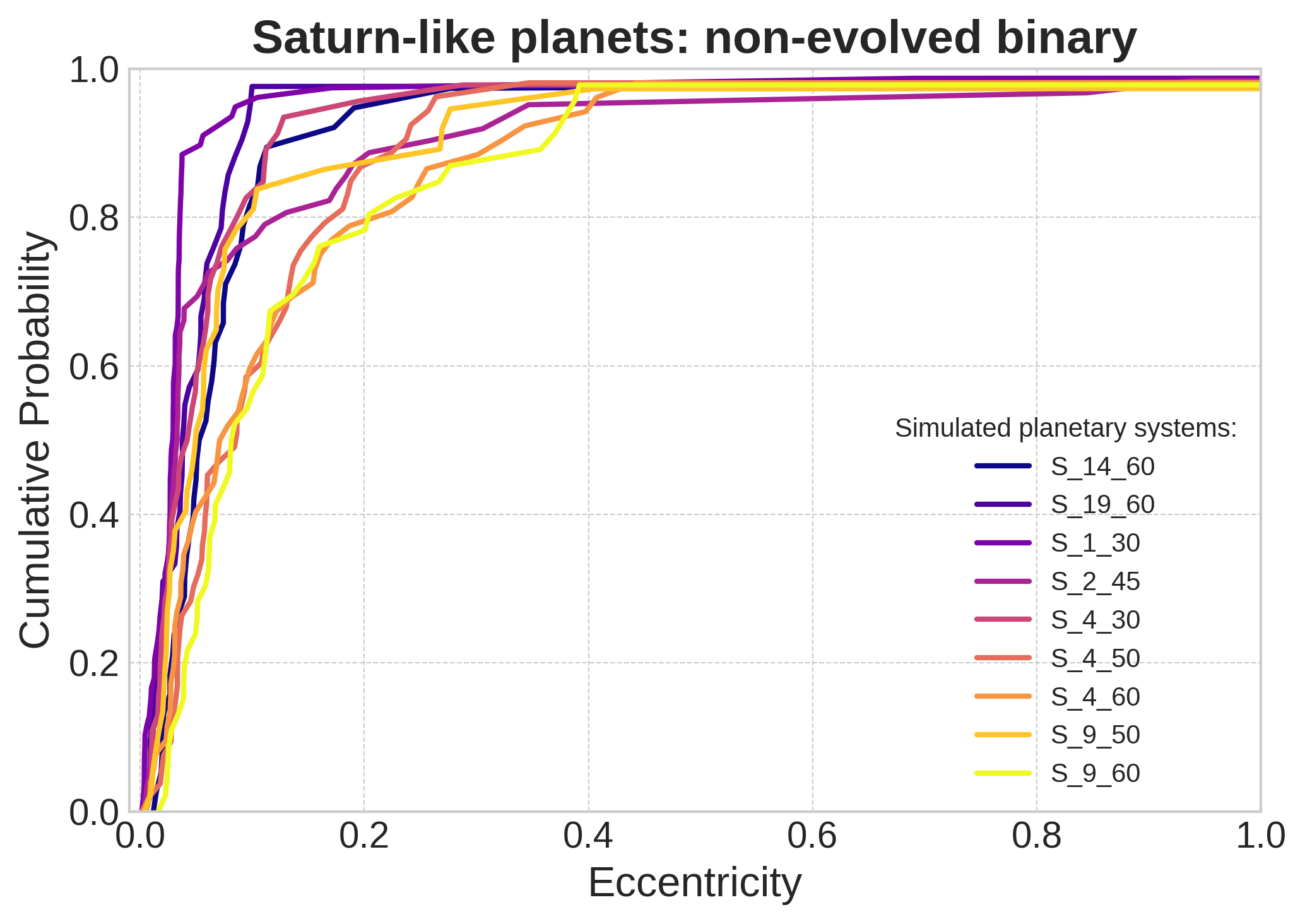}
     \includegraphics[width=\columnwidth]{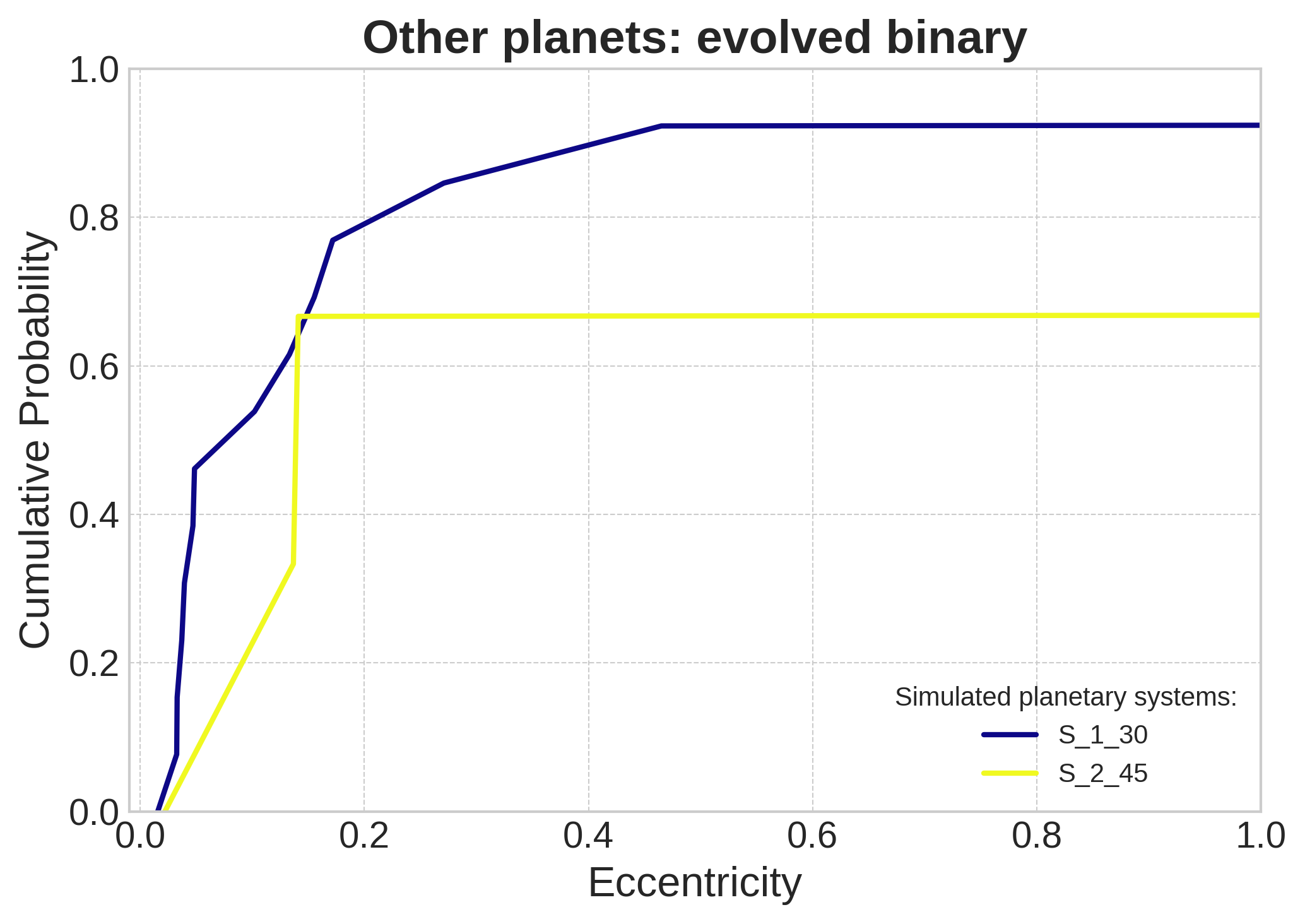}
     \includegraphics[width=\columnwidth]{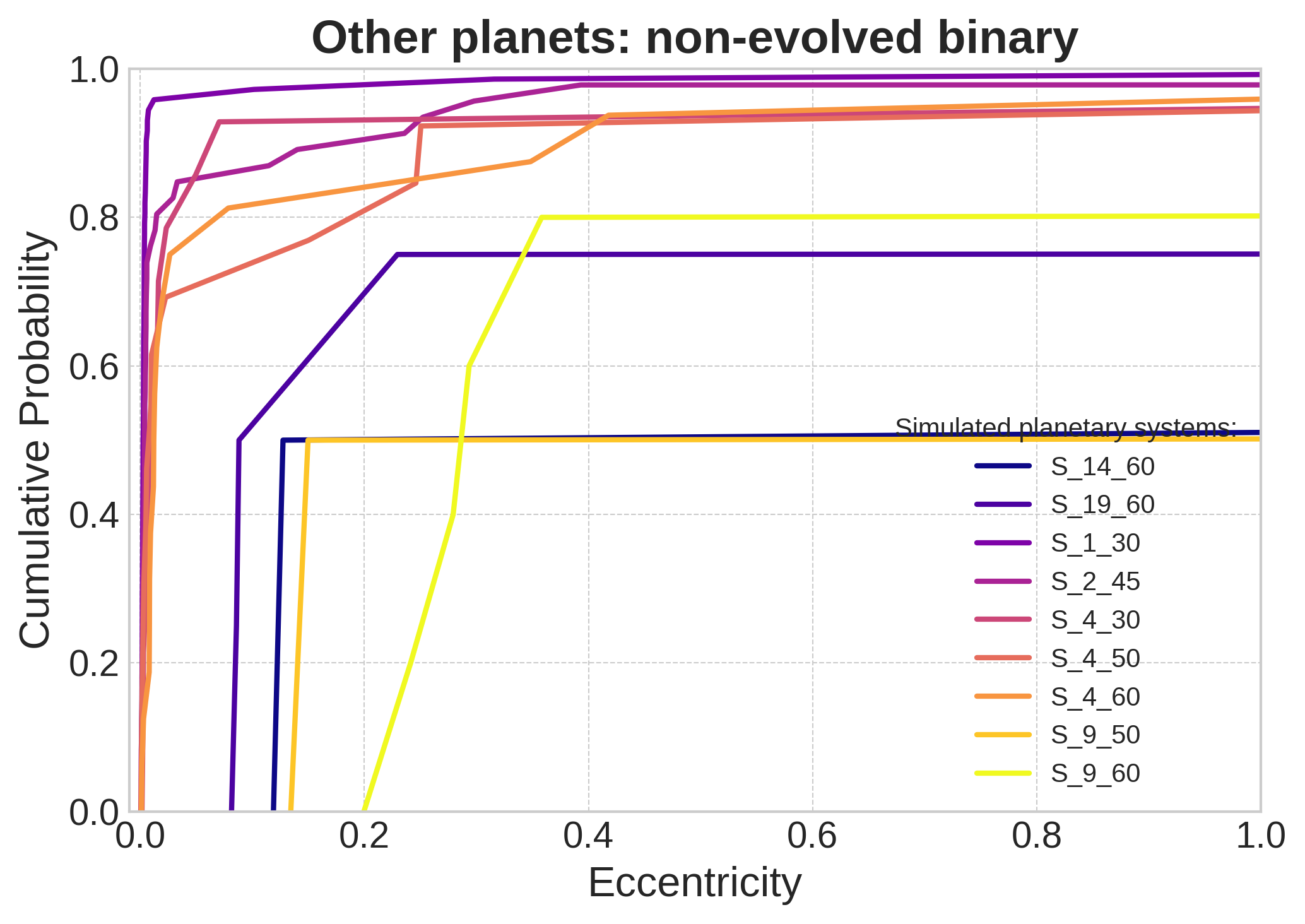}
\caption{
Cumulative normalized distributions of eccentricity for Jupiter-like, Saturn-like, and other planets simulated for different planetary systems (see Table~\ref{tab:simulation_mass}). The left panels show the evolved binary simulations and the right panels show the control simulations (non-evolved binary).
}
\label{fig:cdf_eccentricity}
\end{figure*}

Figure~\ref{fig:cdf_eccentricity} shows the cumulative probability distributions of the final eccentricities of the surviving Jupiter-like planets, Saturn-like planets, and lower-mass planets. The simulations with stellar evolution and the control group show that the eccentricities of  the surviving planets remain below $e\simeq0.4$. Planets with stronger eccentricity excitation ($e\gtrsim0.5$) are usually removed from the system through collisions or ejections; therefore, these planets do not survive until the end of the simulations. 
For Jupiter-like planets, the control runs exhibit these moderately eccentric distributions, with most planets occupying nearly circular or mildly eccentric orbits. When stellar evolution is included, the distributions become slightly broader, reflecting the combined effects of the stellar mass loss and late-stage dynamical interactions. Nevertheless, even in these cases, the surviving planets do not reach eccentricities significantly above $e\simeq 0.3$, highlighting a clear survival threshold in eccentricity space.

The Saturn-like planets exhibit dynamically excited eccentricity distributions during the early stages of the simulations, particularly in the control runs, where planet--planet scattering can temporarily increase eccentricities. However, in both the control simulations and those that include stellar evolution, the final eccentricities of the surviving Saturn-like planets remain below $e\simeq0.15$. Saturn-like planets that experience stronger eccentricity excitation ($e\gtrsim0.15$) are typically removed from the system by collisions or ejections. When stellar evolution is included, the eccentricity distributions become dynamically hotter at intermediate stages, reflecting the amplification of scattering processes during and after stellar mass loss. Nevertheless, despite this enhanced excitation, only Saturn-like planets that maintain eccentricities below $e\simeq0.15$ persist on stable long-term orbits, indicating a more restrictive eccentricity survival threshold compared to Jupiter-mass planets.

In the control simulations, the eccentricity distributions of other planets than Jupiter-like and Saturn-like planets show that the vast majority of planets maintain nearly circular orbits, especially in planetary systems with low exoplanets (see the hot colors in the right panel of Figure~\ref{fig:cdf_eccentricity}). This simulation features multiple exoplanets with eccentricities greater than 0.3. 
When stellar evolution is included, the only two surviving simulations show distributions become moderately high eccentricities of approximately ${0.3-0.6}$. Therefore, stellar evolution significantly increases planetary eccentricities compared to the non-evolved control group.

\subsubsection{Cumulative distribution of the inclination}
\label{CDF_inclination}

\begin{figure*}[ht!]
    \includegraphics[width=\columnwidth]{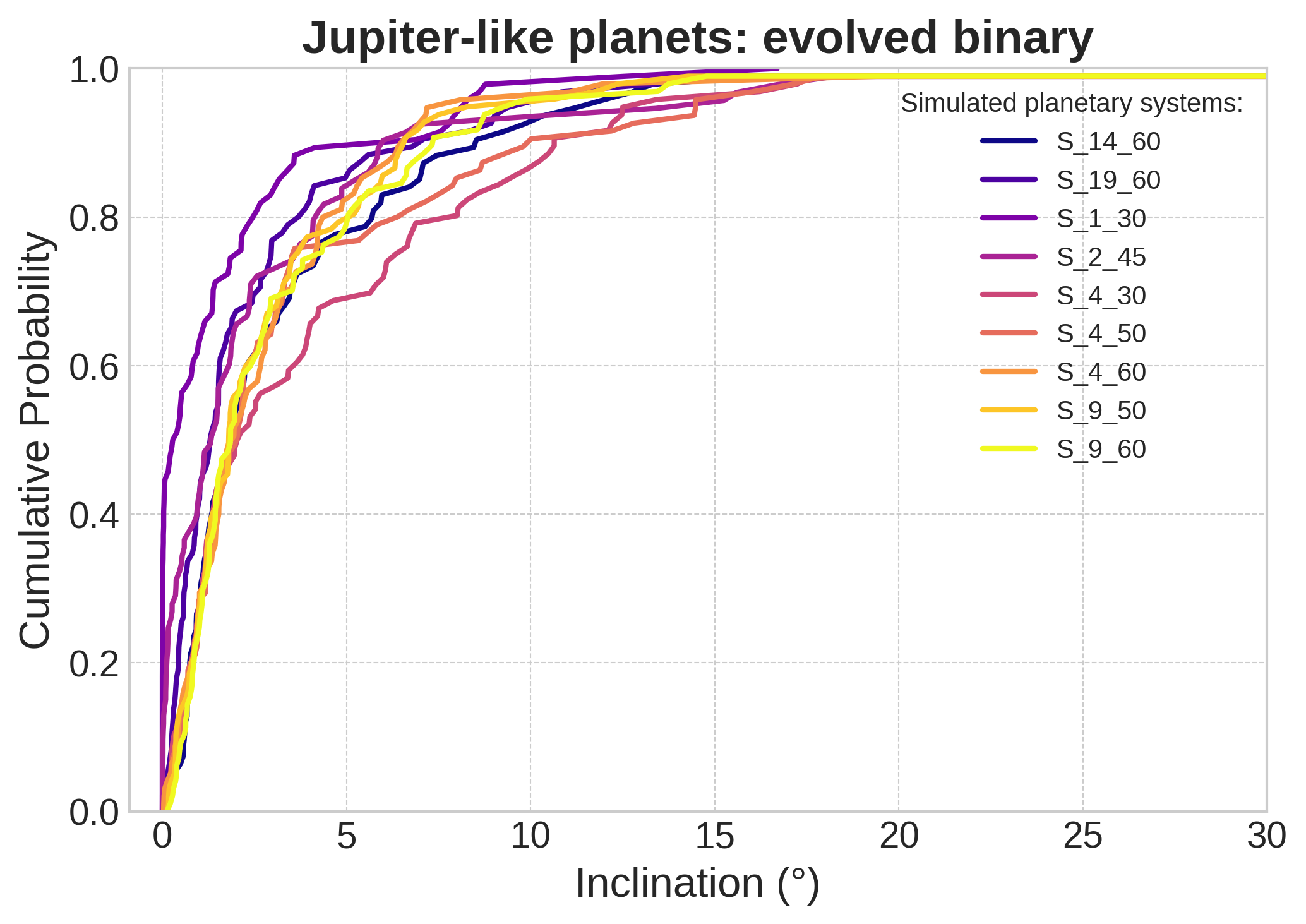}
    \includegraphics[width=\columnwidth]{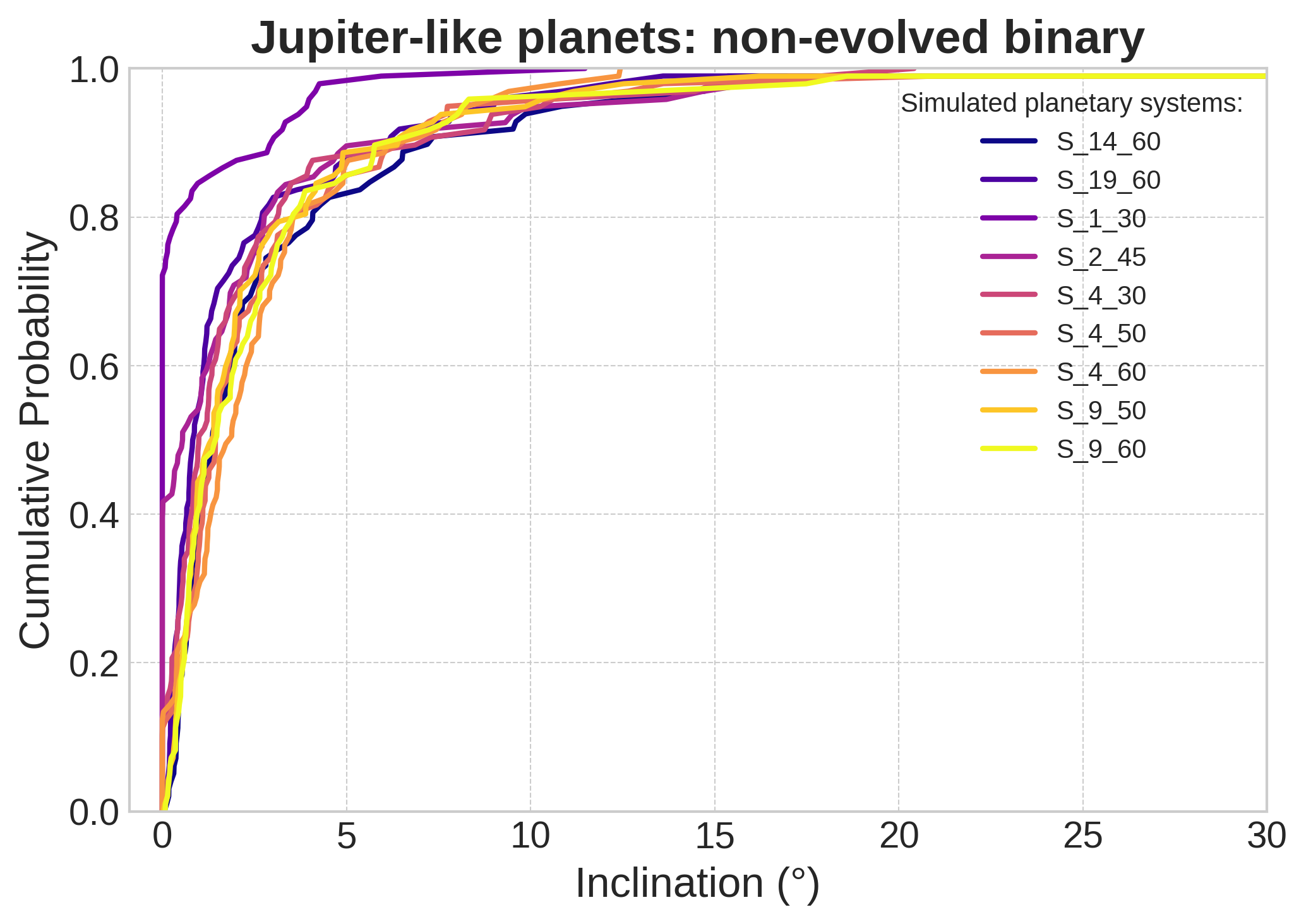}
    \includegraphics[width=\columnwidth]{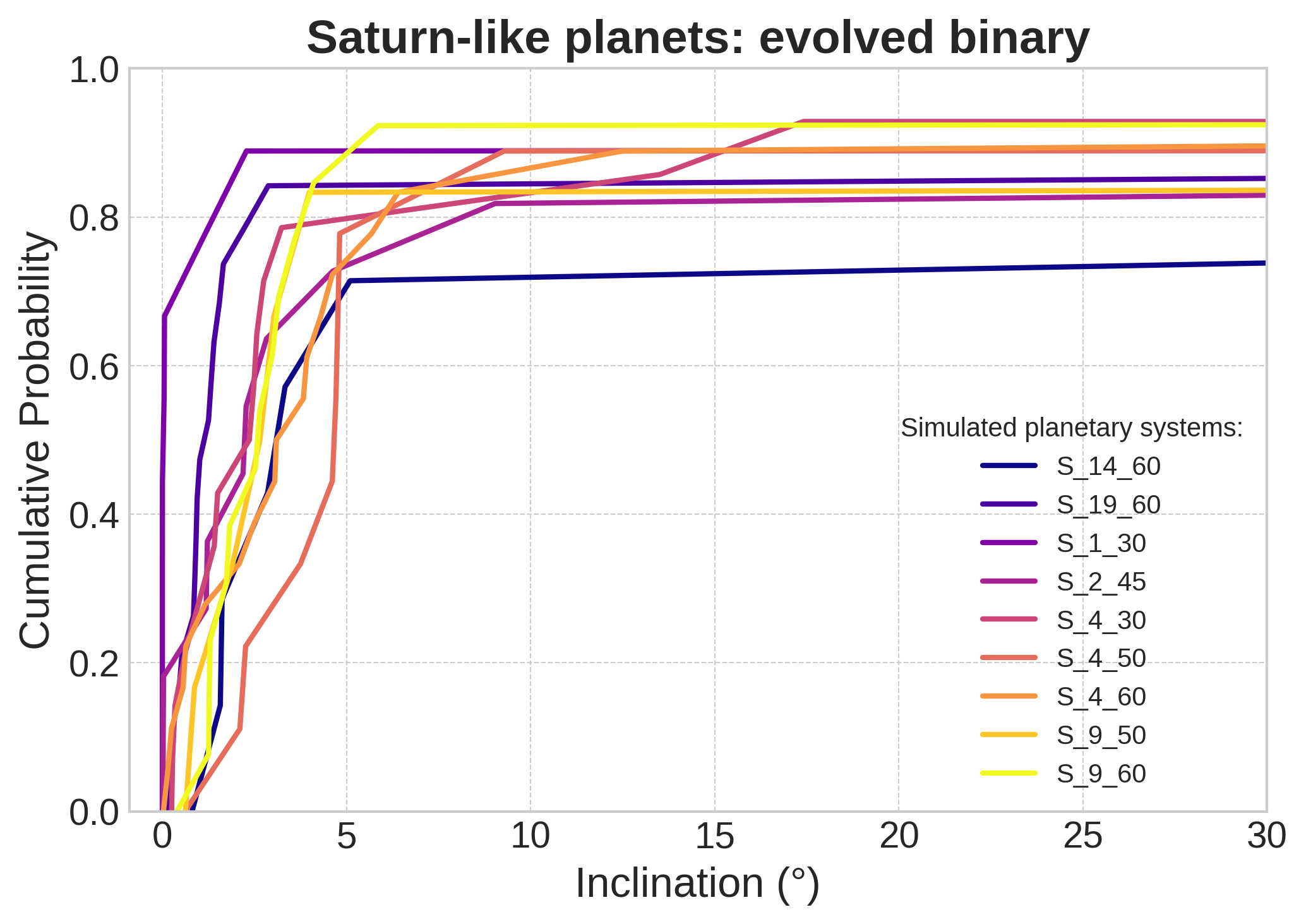}
    \includegraphics[width=\columnwidth]{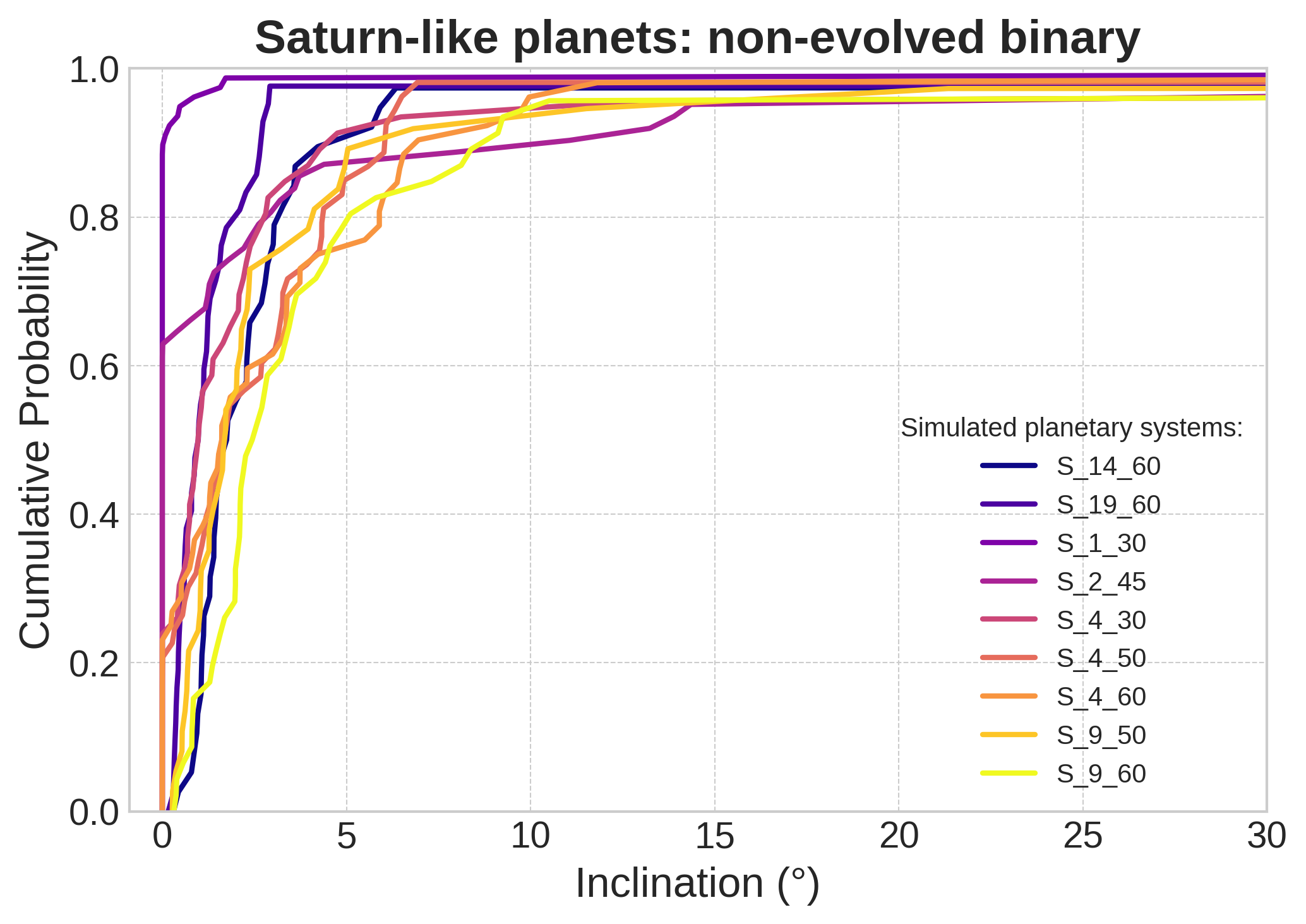}
    \includegraphics[width=\columnwidth]{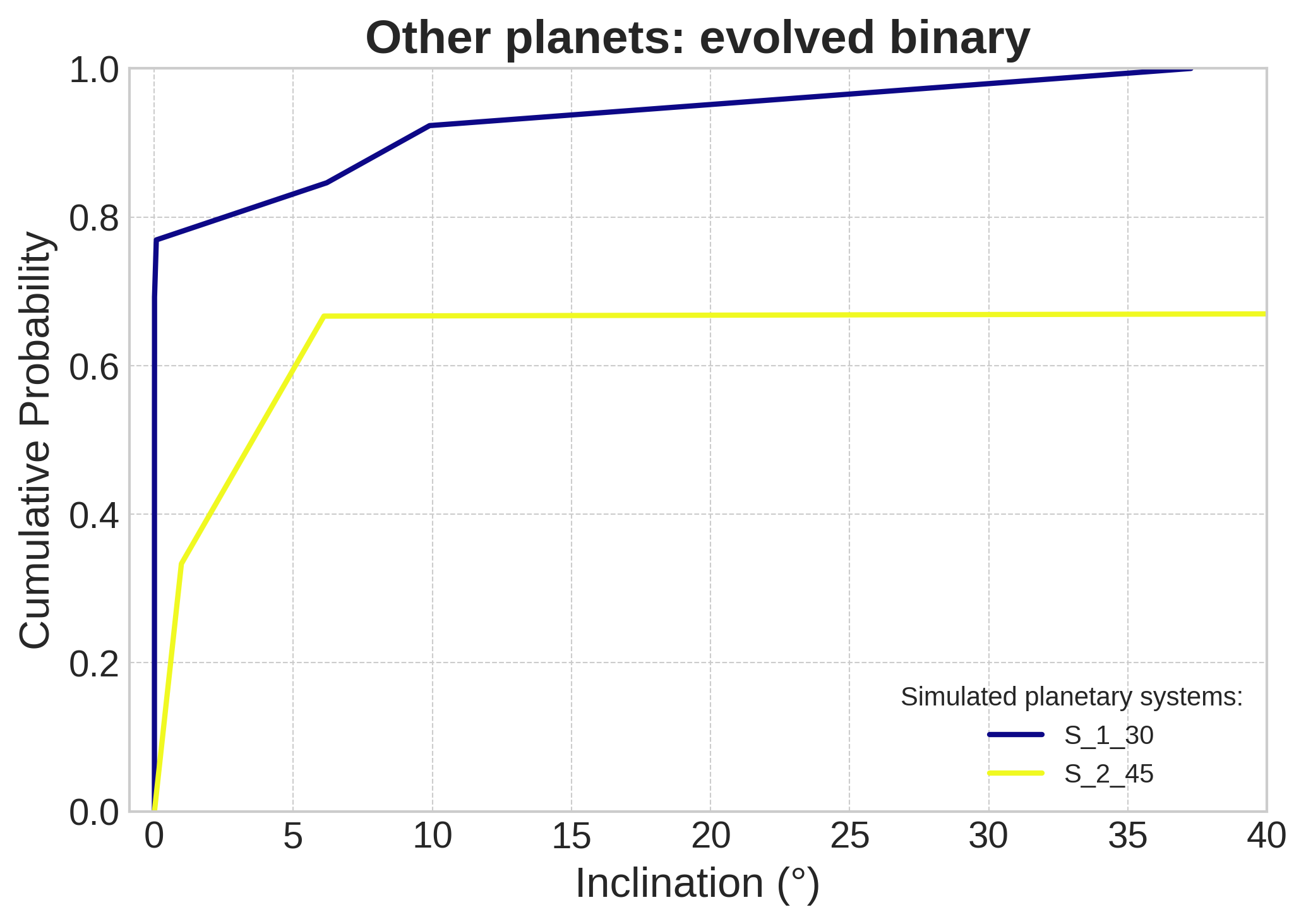}
    \includegraphics[width=\columnwidth]{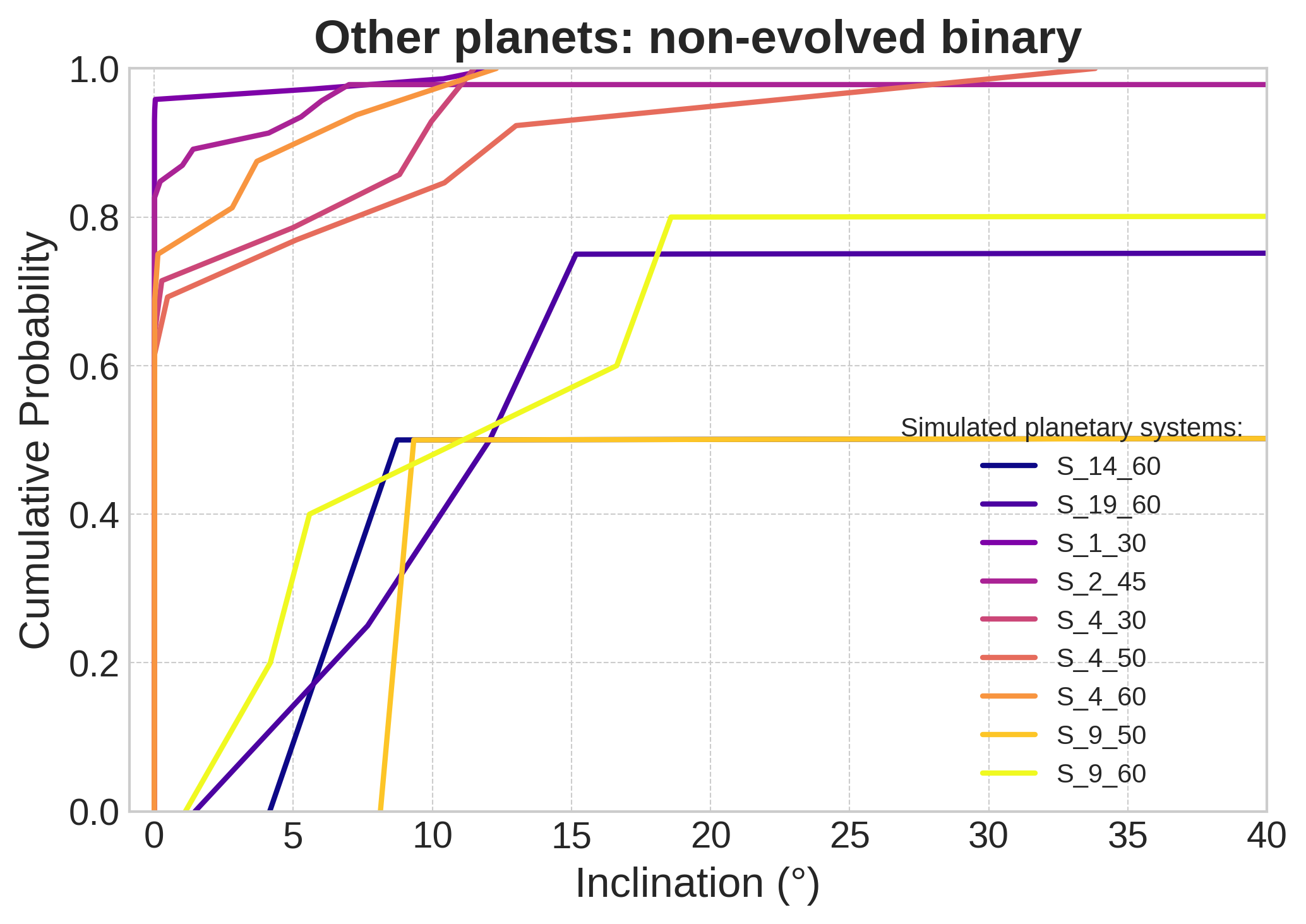}
\caption{
Cumulative normalized distributions of inclination for Jupiter-like, Saturn-like, and other planets simulated for different planetary systems (see Table~\ref{tab:simulation_mass}). The left panels show the evolved binary simulations and the right panels show the control simulations (non-evolved binary).
}
\label{fig:cdf_inclination}
\end{figure*}

Figure~\ref{fig:cdf_inclination} shows the cumulative probability distributions for the final inclinations of the surviving Jupiter-like planets, Saturn-like planets, and lower-mass planets. The inclination distributions of Jupiter-like planets for evolved binary and control simulations are dynamically cold, with more than 80\% of the planets, in the case of evolving system, and more than 90\% of the planets of non evolving systems, with inclinations $i~=~5^\circ-10^\circ$, which establishes a limit value for the inclination. Jupiter-like planets remain largely resistant to strong inclination excitation.

The inclination distributions of surviving Saturn-like planets in control simulations are more inclined than their Jupiter-mass counterparts, with a non-negligible fraction reaching $i\sim 5^\circ$~-~15$^\circ$. This reflects their greater susceptibility to inclination excitation during scattering events. With stellar evolution simulations, the inclination distributions become systematically broader. A larger fraction of Saturn-like planets are driven to intermediate and high inclinations ($i\sim5^\circ~-~25^\circ$), consistent with enhanced dynamical excitation during the post-main-sequence phase.

Figure~\ref{fig:cdf_inclination} also presents the inclination distributions of planets other than Jupiter-like and Saturn-like. In the control simulations, these planets exhibit significant inclination excitation, with many of them reach $i \sim 10^\circ-20^\circ$ for a population with more exoplanets. In simulations with 2 and 3 exoplanets with 15~$M_{\oplus}$, the inclusion of stellar evolution further broadens the distributions, shifting the cumulative probability towards higher inclinations and producing a vertically extended population with inclinations up to $\sim$30$^\circ$.

\added{\subsection{The role of the secondary star}} \label{sec_secondary}

\added{The properties of exoplanets in binaries are different from those of planets orbiting single stars in terms of the period-mass relation. \cite{Zucker_2002} pointed out that more massive short-period exoplanets are mostly found in binaries, this was corroborated by a statistical analysis in \cite{Eggenberger_2004}. In wide binaries, \cite{Desidera_2007} estimated that the secondary stars with semi-major axes ranging from 100 to 300~au are likely to play a role in the formation and evolution of planetary companions. The authors proposed that the properties of exoplanets orbiting wide binary stars are similar to those of planets orbiting single stars, except that there may be a greater abundance of high-eccentricity planets.}

\begin{figure*}
\gridline{
\fig{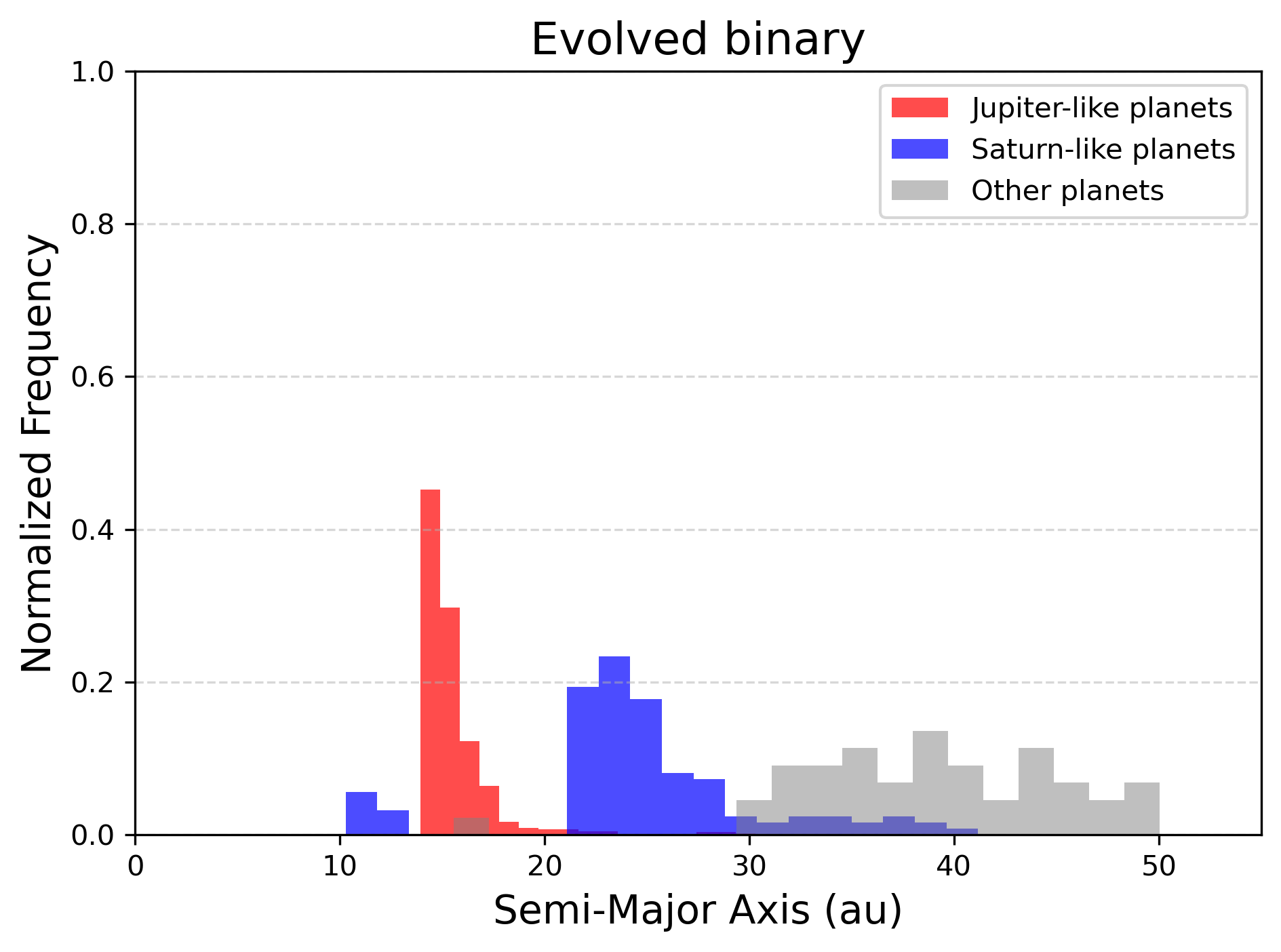}{0.5\textwidth}{(a)} 
\fig{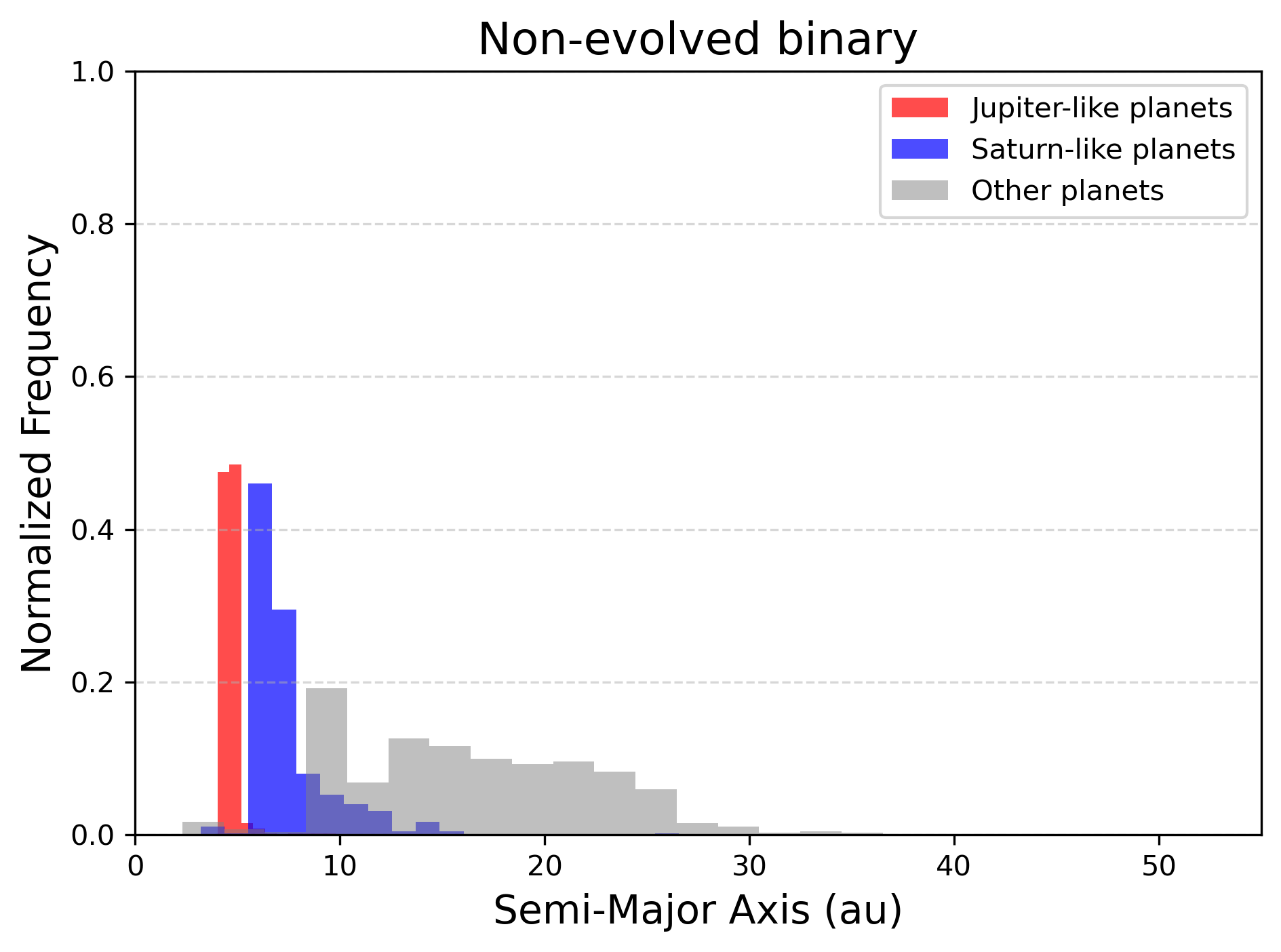}{0.5\textwidth}{(b)}
}
\gridline{
\fig{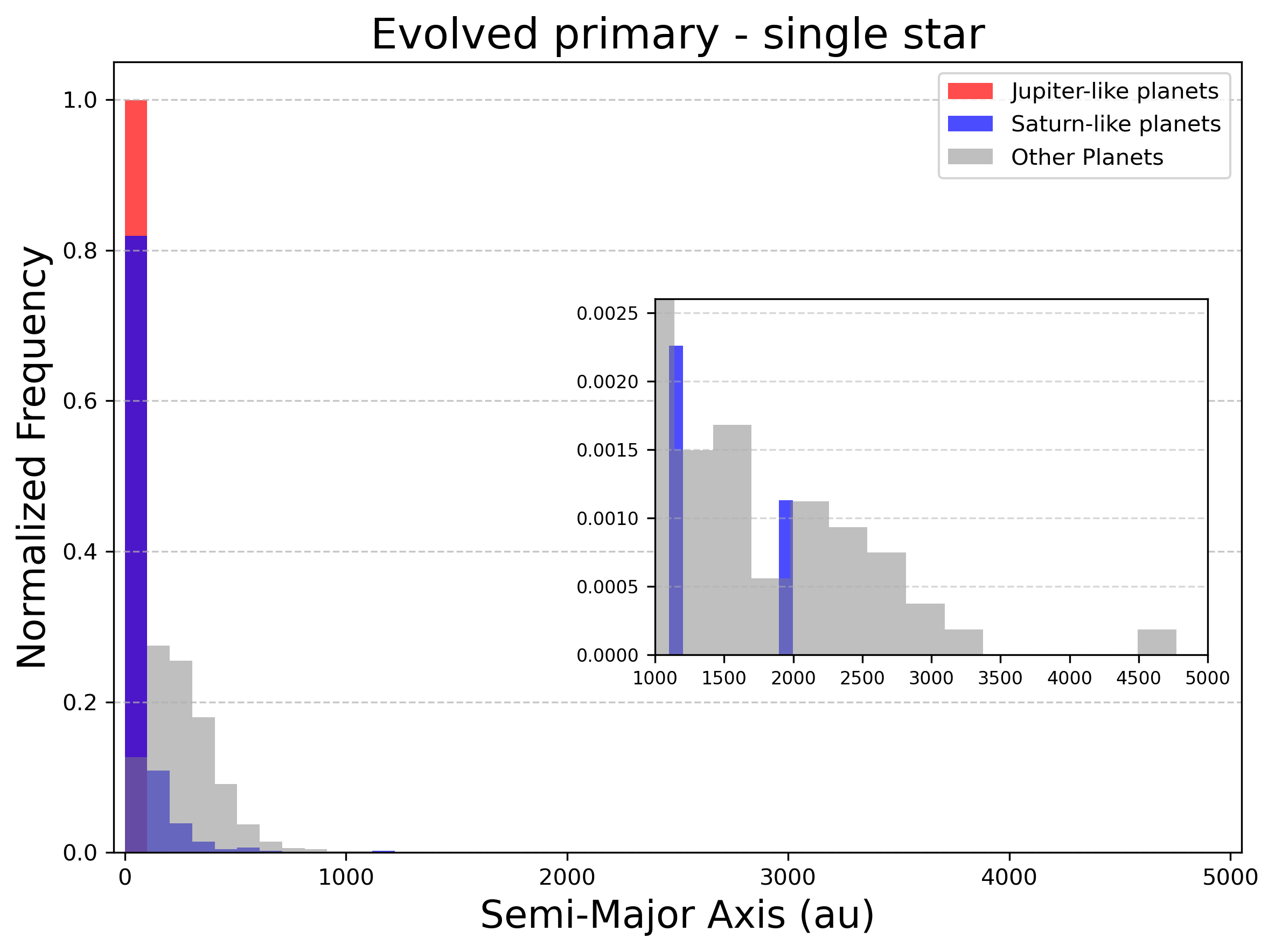}{0.5\textwidth}{(c)} 
\fig{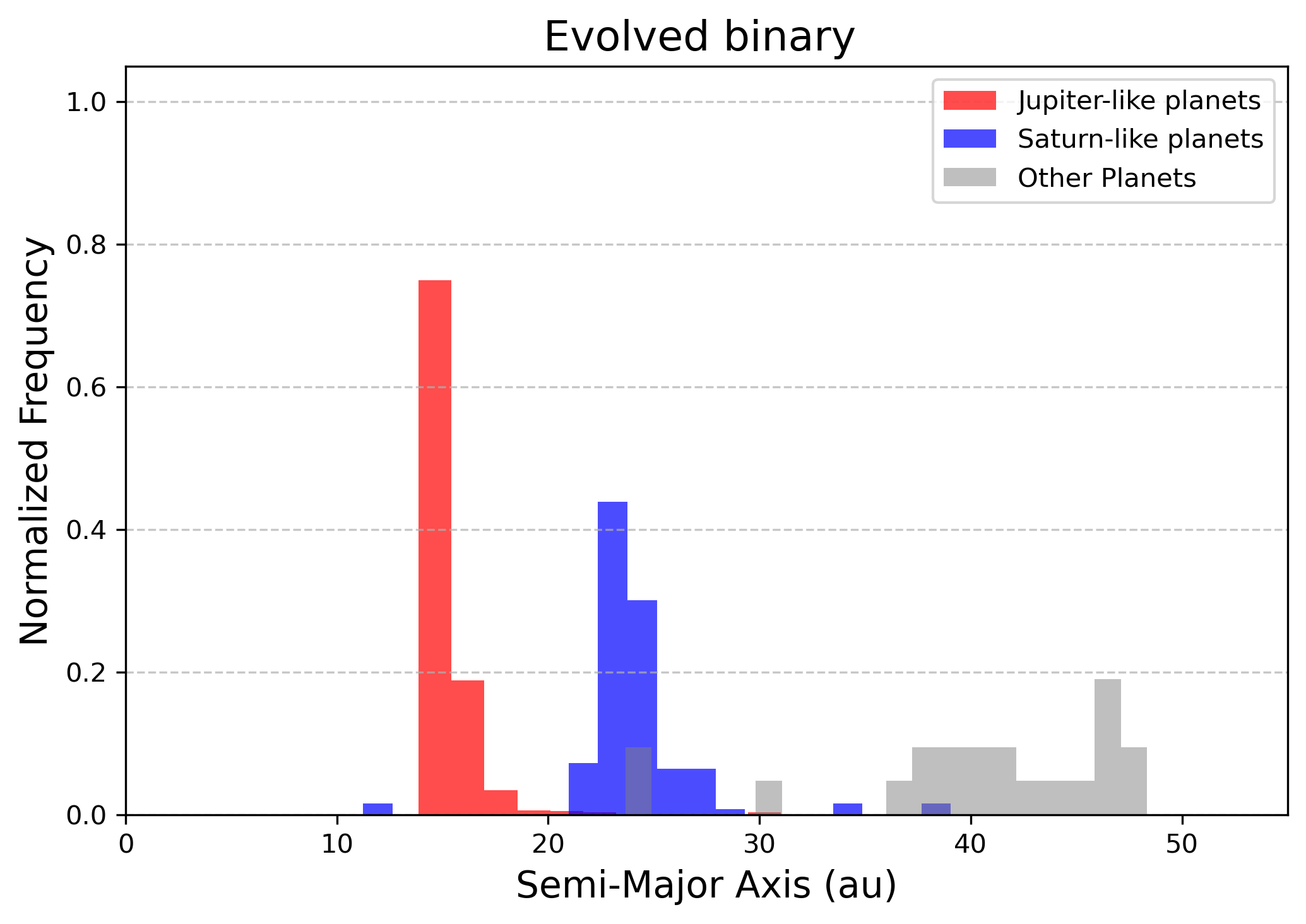}{0.53\textwidth}{(d)}
}
\caption{Histograms showing the distributions of the semi-major axis of the surviving exoplanets. The color red represents Jupiter-like planet, the color blue represents Saturn-like planet, and color gray represents other planets with masses similar to Neptune and Uranus. Panel (a) shows a simulation of 8,400 exoplanets in various simulated planetary configurations as detailed in Table~\ref{tab:simulation_mass} for an evolved binary system. Panel (b) shows the same setup as panel (a), but for a non-evolved binary system. Panel (c) shows a simulation of 1,000 samples from Sim. TAG = ${19-60}$ related to an evolved primary in a single-star planetary system containing one Jupiter-like planet, one Saturn-like planet, and 19 other planets with a mass of 3~$M_{\oplus}$. Panel (d) uses the same setup as Panel (c), but considers an evolved binary.}
\label{histogram}
\end{figure*}

\added{To understand the role of the secondary star in wide binary systems, we performed extra 1,000 simulations in the absence of a secondary star. We used a planetary configuration initially with more exoplanets (Sim.~TAG = ${19-60}$) and a primary star that evolved from a main sequence star to a white dwarf. We also simulated a control account comprising 1,000 samples of Sim. TAG = ${19-60}$ with evolved binary stars. Figure~\ref{histogram} shows the histograms of these simulations in panels (c) and (d), respectively for a primary that evolved in a single-star planetary system and in an evolved binary system. Of the 21,000 simulated planets, the number of surviving Jupiter- and Saturn-like planets, other exoplanets with 3~$M_{\oplus}$, is 7235 in the single-star system and 1126 in the evolved binary system. These results suggest that gravitational perturbations from the companion star affect the stability of exoplanets, influencing the subsequent evolution of less massive and more distant planet from primary.}

\added{The secondary star is also responsible for the concentration of less massive exoplanets in the semi-major axis range of 30$–$40~au, Jupiter-like planets at around 15~au, and Saturn-like planets in the 20$–$30~au range (see Panel (d) in Figure~\ref{histogram}). This behavior is consistent with the analysis of the surviving systems with different planetary configurations proposed in Table~\ref{tab:simulation_mass}, as shown in Panel (a) and discussed in Section~\ref{CDF_semi_major}. The exoplanets belong the single star in evolved process have concentrate semi-major axis values for Jupiter- and Saturn-like planets of order 50~au and some exoplanets with mass of 3~$M_{\oplus}$ very scattered reaching distances of up to 4,000~au (Panel c) and more eccentric orbit as well. We detected more mergers between exoplanets in the single-star simulation than in the binary system, respectively, 19 and 2 cases. This result points directly to how profoundly a binary setup alters the dynamical environment of a forming planetary system.}

\added{In fact, our results agree with those of \cite{Marzari_2005}, who studied the statistical properties of S-type orbits containing two and three Jupiter-like planets in a more compact configuration, with the secondary star fixed at 50~au. The authors found that the binary
companion participated directly in the planet scattering, with a few per cent of the planets being ejected upon colliding with the secondary. In our simulation involving an evolved binary with a semi-major axis ranging from 100 to 202~au and a multi-planetary system, we also found that dynamical instability increased compared to the single-star case and concentrated the exoplanets within a semi-major axis of 50~au during the temporal evolution.}


\vspace{0.5cm}

\section{Comparing numerical simulations and observations} \label{observations}

\begin{figure*}
\gridline{\fig{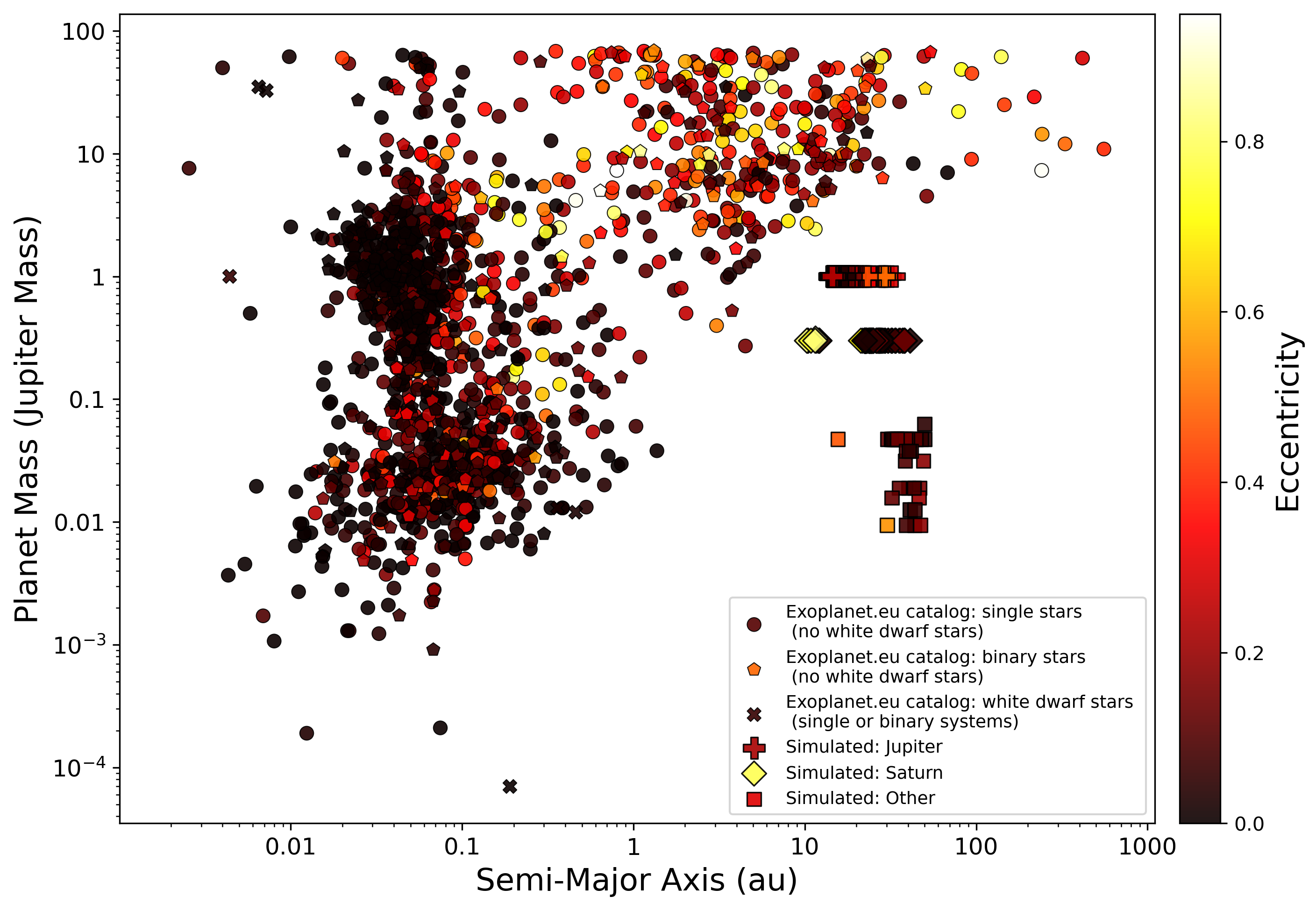}{0.52\textwidth}{(a)}
\fig{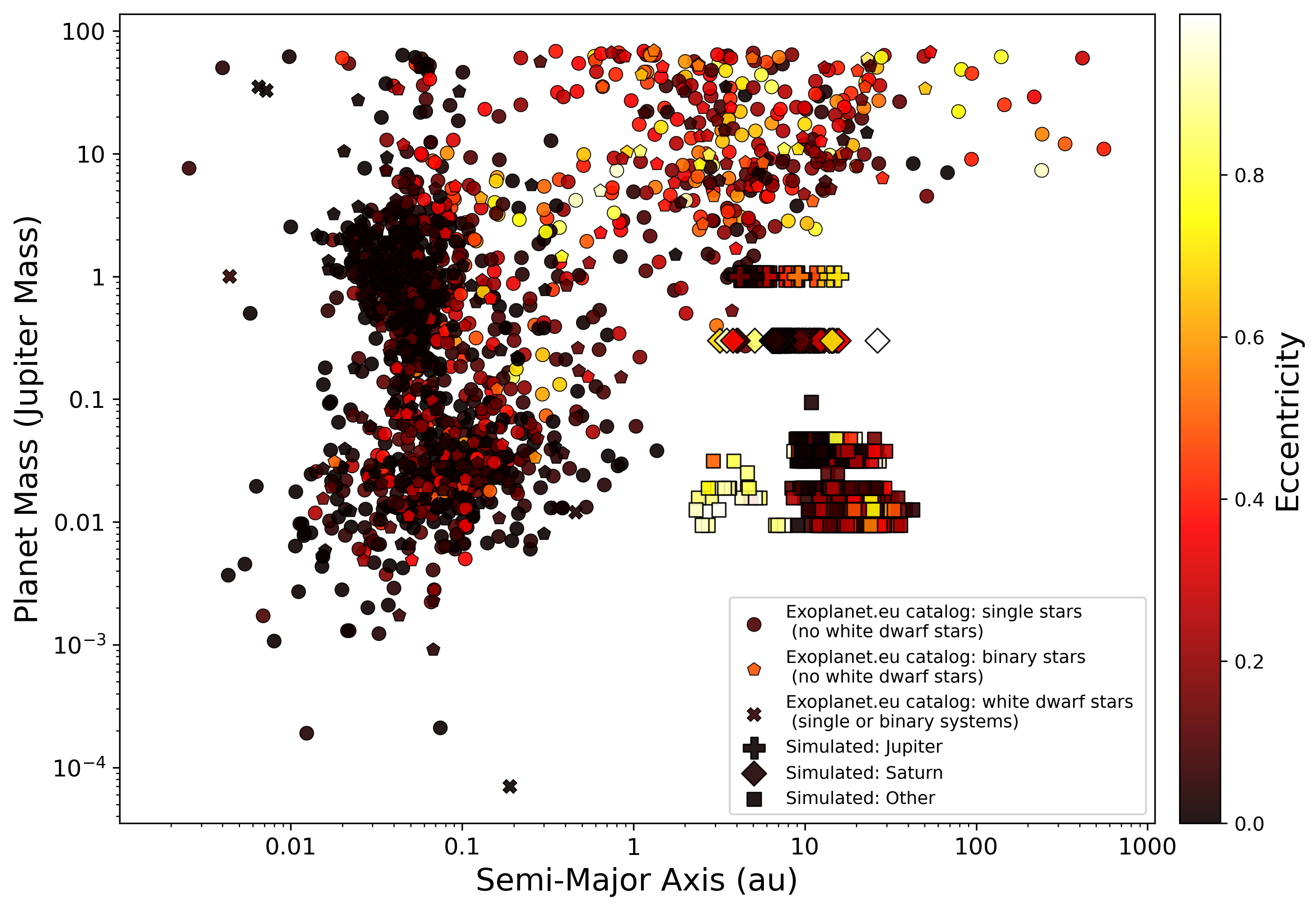}{0.52\textwidth}{(b)}}
\gridline{
\fig{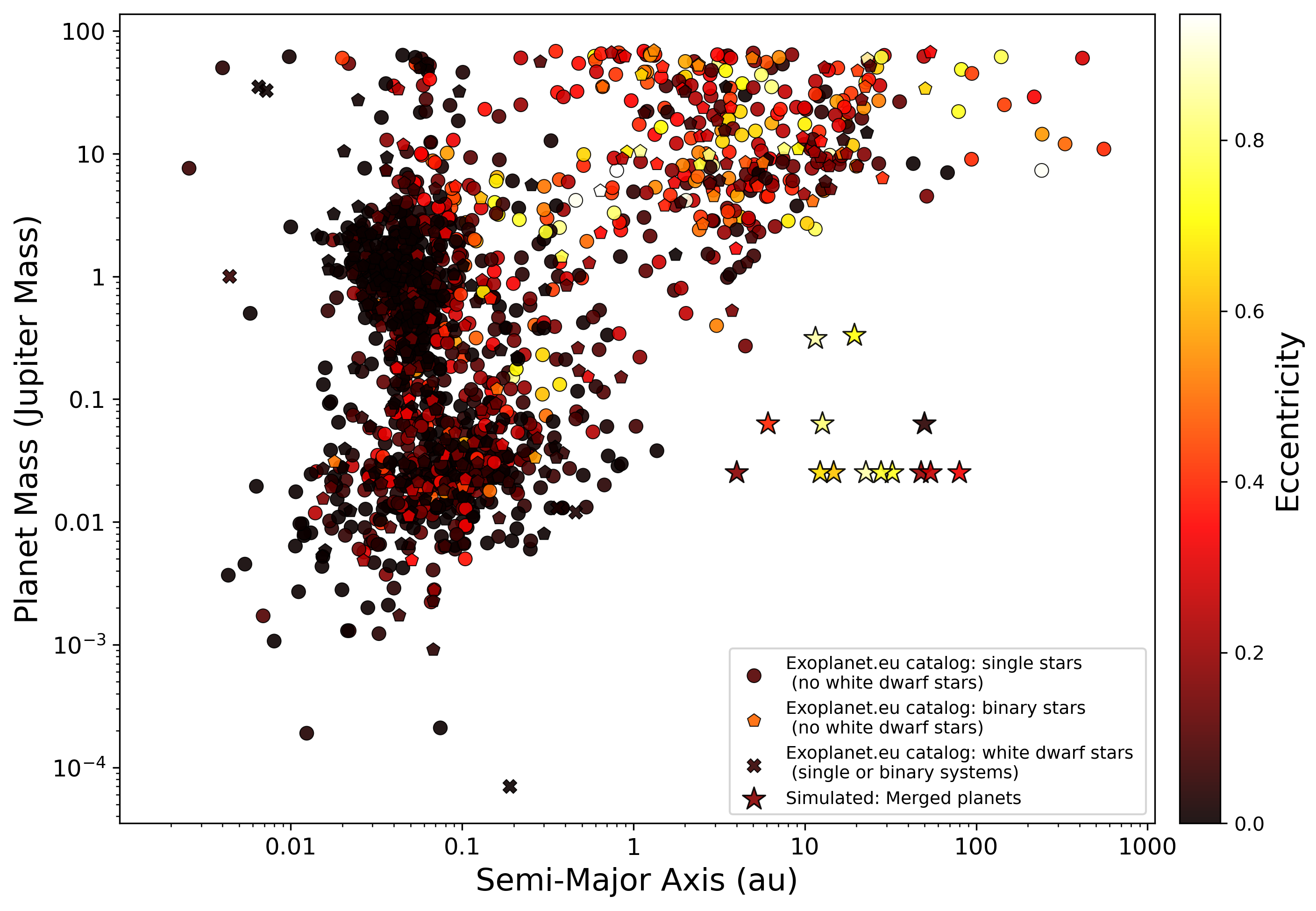}{0.52\textwidth}{(c)} 
\fig{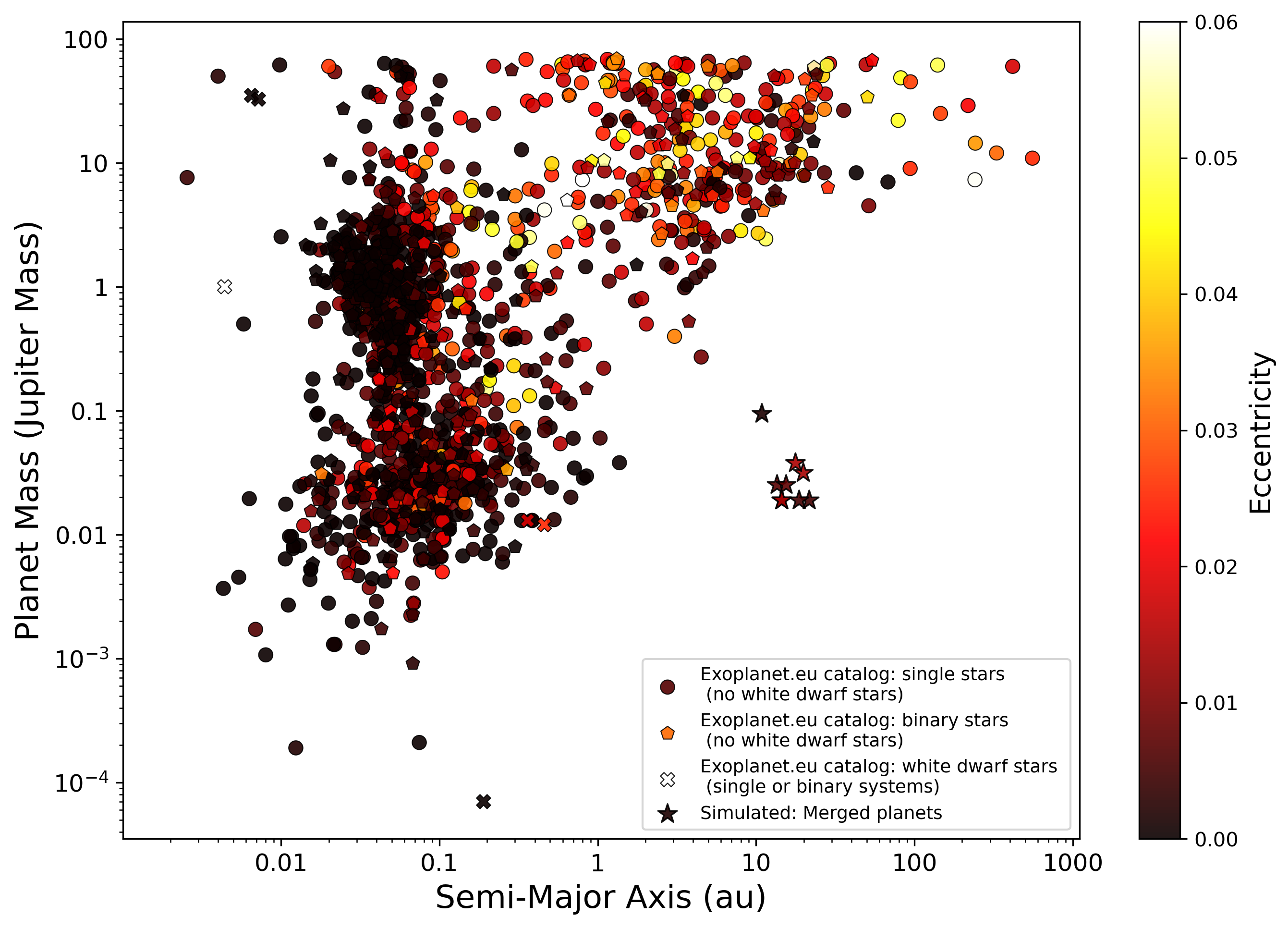}{0.52\textwidth}{(d)}
}
\caption{Currently known exoplanets plotted as a function of semi-major axis and planetary mass. The color coding indicates the orbital eccentricity. Circle symbols indicate exoplanets in single stars with no white dwarf stars, the pentagram symbols indicate exoplanets in binary stars with no white dwarf stars, and the crosses symbols indicate exoplanets in white dwarfs belonging single or binary systems.
First panels show stable planets in evolved binary simulations (left) and in non-evolved binary simulation (right). The symbols of plus, diamond, and square indicate the Jupiter-like planets, Saturn-like planets, and other masses planets, respectively.
Second panels show stable planets obtained by the merging of planets (star symbol) in our simulation with evolved binary systems (left) and non-evolved binary (right).
}
\label{fig:example_simulation_axt1}
\end{figure*}

The current known population of exoplanets (approximately 7,000) has more than three-quarters of  orbital periods  less than 50~days ($\sim$0.3~au). This is the result of observational bias rather than being a feature of the underlying exoplanet population \citep{Mayo_2019}. Figure~\ref{fig:example_simulation_axt1} shows the distribution of these exoplanets (circle symbol is the single stars without white dwarf; pentagon is the binary stars without white dwarf; and `x' is the white dwarf stars in single or binary systems) as a function of semi-major axis and planetary mass, with the color indicating their orbital eccentricity.

A desert region, known as the Neptunian desert, is observable for masses with $0.02 \lesssim M_p \lesssim 0.8~M_{\text{Jup}}$ and orbital periods shorter than 2$-$4~d ($a < 0.1$~au), this region appears as a sparsely populated ``gap" between small rocky super-Earths and massive gas-rich hot Jupiters. A primary explanation for this desert is that planets are highly vulnerable to photoevaporation \cite[e.g.,][]{Owen_2018}. The intense high-energy radiation from the host star can strip away the gaseous envelopes of Neptune-sized planets, leaving behind smaller rocky cores or causing them to migrate out of the region entirely. The other scenario is that Neptune planets are formed and/or evolved differently in different close-in orbital regions \citep[e.g.,][]{Mazeh_2005, Mazeh_2016}. 

Also evident is a clear region of non-exoplanets containing long-period exoplanets with intermediate-sized mass ($0.03 \lesssim M_p \lesssim 0.3~M_{\text{Jup}}$) and Jupiter-like planets. The absence of these exoplanets is due to the observational techniques used in current surveys, rather than physical processes.

The surviving exoplanets obtained from our simulations, both with and without evolution, are concentrated in the uninhabited regions of the semi-major axis and the mass diagram (see Panels (a)–(d) in Figure~\ref{fig:example_simulation_axt1}). 
In panel~(a), which is related to the evolved simulation, \added{874 (of the 1,042 non-eject exoplanets)} of the Jupiter-like survivors (plus symbols) are clustered between 10 and 50~au. 
The points overlap and the eccentricity colors are mixed, but it is possible to see that the eccentricities appear to be moderate to high (as shown by the orange-red colors). The Saturn-like planets (\added{124} of the survivors, represented by diamond symbols) are found at similar orbital distances to the simulated Jupiter-like with $a \approx 3$ to 30~au with moderate to high eccentricities (yellow color).
This suggests that they may be more susceptible to gravitational perturbations or scattering events during the evolution of the system.  
The less massive exoplanets (square symbols), representing only \added{44} of the sample, are distributed across a wide range of semi-major axes in the outer system and with moderate eccentricity.  In the non-evolution simulation (panel b), the Jupiter-like and Saturn-like planets are  
primarily concentrated within semi-major axes of 5~--~30~au, where they are initially positioned at the beginning of the simulation (5 and 6.55~au). The other planets are found between 3 and 30~au, with predominantly dark red or brown eccentricities, indicating relatively low to moderate eccentricities ($e < 0.3$). 
These results suggest that the mass loss from the primary star during its evolution shifts the stable zones for these planets, forcing them to migrate to larger orbital distances in order to maintain stability. The evolution of the stars also indicates an excitation of eccentricity, which pumps the orbits of the surviving planets to moderate values ($e < 0.4$, see Section~\ref{CDF_eccetricity}) probably through secular perturbations, resonance crossings or close encounters.

Panels~(c) and (d) show the surviving exoplanets after collision and merger (star symbols) obtained with and without binary evolution, respectively. In the non-evolved simulation, the merger events are highly localized, occurring within a narrow orbital range of approximately 10 and 20~au. The relatively low eccentricities of these merged bodies ($e < 0.2$) suggest that, in static binary environments, mergers are primarily driven by local dynamical instabilities within a stable orbital belt. 
In contrast, the evolved binary simulations show a dramatic expansion of the merger zone, with events scattered across a much wider range of semi-major axes, extending from 5~au to almost 100~au. 
Furthermore, the merged planets in the evolved scenario exhibit significantly higher and more varied eccentricities (reaching up to $e \approx 0.8$). This indicates that the changing gravitational potential and mass loss associated with stellar evolution trigger large-scale orbital migrations and high-velocity crossings that are absent in the control group. 

The long-period exoplanets obtained in our simulations may be discovered through observation in an era driven by a new generation of space-based observatories and extremely large ground-based telescopes.
Upcoming missions such as the Nancy Grace Roman Space Telescope \citep[e.g.,][]{Mosby_2020}, the PLATO (Planetary Transits and Oscillations of stars) mission \citep{Rauer_2014}, and the Extremely Large Telescope (ELT) \citep{Hook_2010} will provide the statistical census and detailed chemical profiles required to understand the diversity of exoplanetary systems.

\section{Conclusions} \label{conclusions}

In this work, we investigated the dynamical evolution of multi-planetary systems in \added{S-type orbit} of a wide binary system, in which the primary star evolves from the main sequence to the white dwarf stage. \added{We also simulated an evolved single star in a multi-planetary system.}
By coupling \texttt{MESA} stellar evolution tracks with \texttt{REBOUND} N-body integrations, we analyzed the impact of stellar mass loss on planetary stability and orbital architecture. Our results show that stellar evolution \added{in binary system} significantly increases the instability of exoplanetary systems. 
The ejection rates in evolved binary numerical simulations were consistently higher in all configurations compared to the non-evolved control group.
We obtained that Jupiter-like planets are the most dynamically robust population, remaining relatively stable, despite being scattered to wider orbits with low to moderate eccentricity and inclination. Saturn-like planets occupy an intermediate regime and exhibit substantial sensitivity to both planetary scattering and stellar evolution. 
Exoplanets that are less massive than Saturn-like planets are less likely to survive, resulting in statistics that are less comparable to those of other planets. These exoplanets exhibit broad distributions in semi-major axis, eccentricity, and inclination, particularly when stellar mass loss is included. \added{The presence of a secondary star concentrated the less massive survivors within a semi-major axis of 50~au.} In general, stellar evolution \added{in binary systems} plays a fundamental role in shaping the final orbital architecture of planetary systems, with its impact increasing systematically toward lower planetary masses. The orbital parameters obtained in our simulations provide a potential dynamical explanation for the population of exoplanets in the long-period giant region, a regime in which only a few exoplanets are known. Our results indicate that long-period giant exoplanets may be a direct consequence of the coupled effects of binary dynamics and post-main-sequence stellar mass loss.

\begin{acknowledgments}
IJL acknowledges \textit{São Paulo Research Foundation} (FAPESP) for financial support under grant \#2024/14358-9 and \#2024/03736-2 and the Conselho Nacional de Desenvolvimento Científico e Tecnológico (CNPq, \#400990/2022-9 and \#300612/2025-7). RR thanks to the scholarship granted from the Brazilian Federal Agency for Support and Evaluation of Graduate Education (CAPES), in the scope of the Program CAPES-PrInt (Proc~88887.310463/2018-00, Mobility number 88887.572647/2020-00, 88887.468205/2019-00). This research was supported in part by the São Paulo Research Foundation (FAPESP) through the computational resources provided by the Center for Scientific Computing (NCC/GridUNESP) of the São Paulo State University (UNESP). RR also acknowledges support provided by grants FAPESP (Proc~2016/24561-0), by São Paulo State University (PROPe~13/2022) and (Proc~2022/11783-5). RR also thanks to CNPQ 405349/2025-4. DB is supported by the National Natural Science Foundation of China (NSFC, Nos. 12288102, 12125303), the Strategic Priority Research Program of the Chinese Academy of Sciences (grant Nos. XDB1160303, XDB1160300, XDB1160000, XDB1160200, XDB1160201), the CAS Project for Young Scientists in Basic Research (YSBR-148), the Yunnan Revitalization Talent Support Program ``YunLing Scholar'' project, International Centre of Supernovae (ICESUN), Yunnan Key Laboratory of Supernova Research (No. 202505AV340004),  the New Cornerstone Science Foundation through the XPLORER PRIZE, and the Yunnan Revitalization Talent Support Program-Science \& Technology Champion Project (No. 202305AB350003). SGW thanks FAPESP (Proc~2022/11783-5) and CNPq (Proc.~309057/2025-6) for the financial support. E.M. acknowledges funding from FAPEMIG under project number APQ-02493-22 and a research productivity grant number 309829/2022-4 awarded by the CNPq. RANA thanks CNPq (No. 405349/2025-4).
\end{acknowledgments}

\facilities{}

\software{MESA \citep{Paxton2011,Paxton2013,Paxton2015,Paxton2018,Paxton2019,Jermyn2023},  
          REBOUND \citep{Rein2012, Rein2015Spiegel, Rein2015Tamayo}.
          }




\bibliography{sample701}{}
\bibliographystyle{aasjournalv7}



\end{document}